\documentclass[preprint,comments,revision]{ptephy_om}
\preprintnumber{RIKEN-iTHEMS-Report-26}
\revisionnum{1}

\renewcommand{\revone}[1]{#1}

\usepackage{tikz}
\usepackage{booktabs}

\usepackage{comment}

\usepackage{hyperref}
\usepackage{orcidlink}

\title{Complex scaling approach to quasinormal modes of Schwarzschild and Reissner--Nordstr\"om black holes}

\author{Shoya Ogawa \orcidlink{0000-0003-0900-2486}}
\affil{Department of Physics, Kyushu University, 744 Motooka, Nishi-ku,
Fukuoka 819-0395, Japan
\email{ogawa.shoya.615@m.kyushu-u.ac.jp}
}

\author[2]{Takuya Hirose \orcidlink{0000-0003-0962-8884}}
\affil[2]{Faculty of Science and Engineering, Kyushu Sangyo University,
Fukuoka 813-8503, Japan
\email{t.hirose@ip.kyusan-u.ac.jp}
}

\author[3]{Okuto Morikawa \orcidlink{0000-0002-0044-4491}}
\affil[3]{Center for Interdisciplinary Theoretical and Mathematical Sciences (iTHEMS),
RIKEN, Wako 351-0198, Japan
\email{okuto.morikawa@riken.jp}
}

\begin{document}
\begin{abstract}
We study black-hole quasinormal modes by applying the complex scaling method
(CSM) to the perturbation equations of Schwarzschild and
Reissner--Nordstr\"om black holes.
The method converts the outgoing-wave boundary condition into a
non-Hermitian eigenvalue problem, allowing quasinormal-mode frequencies to be
computed within a common spectral framework.
We first benchmark the method for the Schwarzschild Regge--Wheeler equation and
then extend it to the Reissner--Nordstr\"om family, including the extremal
limit.
Our results show that CSM provides a unified and flexible approach to the
computation of black-hole quasinormal frequencies.
\end{abstract}
\maketitle

\tableofcontents

\section{Introduction}\label{sec:introduction}

Black-hole quasinormal modes (QNMs) are damped oscillations that characterize
the linear response of a black hole to external perturbations.
They are defined by imposing purely ingoing boundary conditions at the event
horizon and purely outgoing boundary conditions at spatial infinity.
Because these boundary conditions render the problem non-self-adjoint, the QNM
frequencies are complex, with their real and imaginary parts encoding the
oscillation frequency and damping rate, respectively.
QNMs therefore occupy a central place in black-hole perturbation theory and in
the interpretation of gravitational-wave ringdown signals~\cite{Leaver:1986gd,Ching:1994bd}.

From a mathematical point of view, QNMs are resonances.
Accordingly, their natural characterization is in terms of the analytic
structure of the resolvent or Green function in the complex-frequency plane,
rather than as ordinary normalizable eigenstates.
For Schwarzschild black holes, this viewpoint leads to the familiar
pole--branch-cut decomposition of the retarded Green function:
QNM poles govern the intermediate-time ringdown, while the branch cut is tied
to continuum contributions and late-time tails
\cite{Leaver:1986gd,Ching:1994bd,Casals:2013mpa,Su:2026fvj}.
This perspective is conceptually important for the present work.
In practical computations, one often seeks to recover resonances from a basis
truncation or discretization cast as a generalized eigenvalue problem.
While this can be effective, it may blur the distinction between genuine
resonances and artifacts tied to the chosen representation.
Our aim is therefore not merely to introduce another numerical scheme, but to
develop a formulation in which the passage from resonance poles to discrete
eigenvalues is theoretically controlled.

For the computation of QNM frequencies themselves, Leaver's continued-fraction
method remains one of the most powerful and accurate approaches
\cite{Leaver:1985ax,Leaver:1986gd}.\footnote{%
Traditionally, black-hole QNM frequencies have been studied by a wide
variety of methods; see, for example, the reviews~\cite{Kokkotas:1999bd, Berti:2009kk,Konoplya:2011qq} and
references therein.}
Its success, however, relies strongly on the special analytic structure of the
radial equation, in particular on the availability of suitable Frobenius-type
expansions and recurrence relations.
This is not just a technical detail.
As one moves away from the standard Schwarzschild problem, the analytic
machinery underlying the continued-fraction construction becomes increasingly
fragile.
For nonextremal Reissner--Nordstr\"om black holes, one can still formulate a
Leaver-type treatment, but the situation changes qualitatively in the extremal
limit, where the two horizons merge and the naive Frobenius structure at the
outer horizon is lost.
This difficulty has motivated dedicated studies of extremal
Reissner--Nordstr\"om QNMs and their continued-fraction treatment
\cite{Onozawa:1995vu,Richartz:2015saa,Daghigh:2024wcl}.

At the same time, recent years have seen a variety of new developments in the
analysis of black-hole QNMs.
These include reformulations based on hyperboloidal slicing and spectral
decompositions, which make the boundary behavior more regular and enable
controlled mode expansions
\cite{Ansorg:2016ztf,Jaramillo:2020tuu,PanossoMacedo:2024nkw},
as well as new analytic directions such as exact WKB approaches to black-hole
quasinormal spectra \cite{Miyachi:2025ptm}.
More algorithmic developments have also appeared for more complicated
perturbation problems, including flexible spectral and hybrid numerical
frameworks \cite{Pombo:2025urp}.
Recent developments have also highlighted resonance phenomena associated with
avoided crossings and exceptional points in black-hole ringdown \cite{Motohashi:2024fwt,PanossoMacedo:2025xnf,Takahashi:2025uwo}.
Taken together, these works make it clear that the field is developing along a
number of complementary directions.
For the present paper, however, our emphasis is different:
we seek a formulation that stays as close as possible to the resonance-theoretic
and scattering-theoretic structure of the problem itself.

The route we pursue is the complex scaling method (CSM).
\revone{Related complex-coordinate ideas were previously used by
Vanzo and Zerbini~\cite{Vanzo:2004fy}
in the asymptotic analysis of QNMs for multi-horizon black holes.
Our focus here is different: we develop a finite-basis CSM framework
for extracting low-lying QNM poles of Schwarzschild and
Reissner--Nordstr\"om black holes, including the extremal limit.}

Originally developed in resonance problems in atomic, molecular~\cite{Reinhardt:1982,Moiseyev:1979,Herbst:1981ew,Chu:1990,Telnov:2013}, and nuclear
physics~\cite{Myo:2014ypa,Myo:2023heu,Guo:2010zzm,Matsumoto:2010mi,Kruppa:2013ala}, complex scaling reveals resonances as discrete eigenvalues of a
non-Hermitian operator by rotating the asymptotic coordinate into the complex
plane, while rotating the continuum away from the resonance sector
\cite{Aguilar:1971ve,Balslev:1971vb,Simon1972BalslevCombes,
Simon1979ECS,Moiseyev:1998gjp,Myo:2014ypa}.
For the present problem, the significance of CSM is therefore not merely
computational.
Its importance is that it provides a framework in which the resonance problem
itself can be represented as an eigenvalue problem in a mathematically
meaningful way.
In this sense, CSM sheds light on the very point where naive
eigenvalue-based discretizations are most vulnerable:
it offers a controlled representation of outgoing resonances as discrete
spectral data of a complex-scaled operator, rather than treating the QNM
boundary condition as an external numerical constraint imposed after the fact.

The goal of this paper is to apply this viewpoint to black-hole perturbations
in a systematic way.
Concretely, we develop a CSM-based framework for computing QNM frequencies for
the Schwarzschild and Reissner--Nordstr\"om families within a common spectral
formulation.
Schwarzschild serves as the essential benchmark, where the Regge--Wheeler and
Zerilli--Moncrief problems are well understood and where one can compare with
established results based on continued fractions and Green-function
decompositions.
The real target, however, is the extension to charged black holes, including
the approach to the extremal Reissner--Nordstr\"om limit, where a formulation
less dependent on special Frobenius structures is especially desirable.

The rest of this paper is organized as follows.
In Sec.~\ref{sec:qnm}, we review the perturbation equations relevant to
Schwarzschild and Reissner--Nordstr\"om black holes.
In Sec.~\ref{sec:csm}, we summarize the theoretical basis of complex scaling
and formulate the present numerical CSM approach.
In Sec.~\ref{sec:numerics}, we present the resulting QNM spectra for
Schwarzschild and Reissner--Nordstr\"om black holes and discuss their behavior
toward extremality.
Finally, Sec.~\ref{sec:conclusion} summarizes our results and future
perspectives.

\section{Review of Regge--Wheeler and Zerilli--Moncrief equations}\label{sec:qnm}
\subsection{Linearized Einstein equations around a fixed background}

We briefly review the perturbative setup that leads to the $1$-dimensional wave equation used in the complex-scaling analysis.
Let the spacetime metric be decomposed as (we use geometrized units $c=G=1$)
\begin{align}
 g_{\mu\nu} =g^{(0)}_{\mu\nu}+h_{\mu\nu},
\end{align}
where $g^{(0)}_{\mu\nu}$ is the background metric and $h_{\mu\nu}$ is a linear perturbation satisfying $|h_{\mu\nu}|\ll 1$.
To first order in $h_{\mu\nu}$, the Christoffel symbol is decomposed as
\begin{align}
 \Gamma^{\lambda}_{\mu\nu}
 =\Gamma^{\lambda~(0)}_{\mu\nu}+\Gamma^{\lambda~(1)}_{\mu\nu},
\end{align}
with
\begin{align}
 \Gamma^{\lambda~(1)}_{\mu\nu}
 =\frac{1}{2}g^{(0)\lambda\rho}
 \left(
 \nabla_{\mu}h_{\nu\rho}
 +\nabla_{\nu}h_{\mu\rho}
 -\nabla_{\rho}h_{\mu\nu}
 \right),
\end{align}
where $\nabla_{\mu}$ denotes the covariant derivative compatible with the background metric $g^{(0)}_{\mu\nu}$.
The Ricci tensor similarly splits into
\begin{align}
 R_{\mu\nu}=R^{(0)}_{\mu\nu}+R^{(1)}_{\mu\nu},
\end{align}
and for a vacuum background satisfying $R^{(0)}_{\mu\nu}=0$ the linearized Einstein equation is simply
\begin{align}
 R^{(1)}_{\mu\nu}=0.
\end{align}
This is the starting point for the Regge--Wheeler and Zerilli formalisms.

\subsection{Odd/even decomposition on Schwarzschild}

For the Schwarzschild background,
\begin{align}
 \rmd s^2
 =-f(r)\rmd t^2+f(r)^{-1}\rmd r^2+r^2(\rmd\theta^2+\sin^2\theta\,\rmd\phi^2),
 \qquad
 f(r)=1-\frac{2M}{r},
 \label{eq:Schwarzschild-metric}
\end{align}
metric perturbations can be expanded in tensor harmonics on~$S^2$~\cite{Regge:1957td,Zerilli:1970wzz}.
Under the parity transformation $(\theta,\phi)\to(\pi-\theta,\phi+\pi)$, these perturbations decompose into two independent sectors:
odd-parity (axial) modes transforming as $(-1)^{l+1}$ and  even-parity (polar) modes transforming as $(-1)^{l}$.
After imposing the Regge--Wheeler gauge, the two sectors decouple.

In the odd-parity sector, one may write the perturbation in terms of two radial-temporal amplitudes, traditionally denoted by $h_0^{l m}(t,r)$ and $h_1^{l m}(t,r)$.
The linearized Einstein equations then reduce to a constraint equation and a dynamical equation.
Introducing the master variable as
\begin{align}
 Q(t,r)=\frac{f(r)}{r}h_1(t,r),
\end{align}
one finds a single second-order wave equation after using the tortoise coordinate $r_*$ defined by
\begin{align}
 r_*=r+2M\ln\left(\frac{r}{2M}-1\right),
 \qquad
 \frac{\rmd}{\rmd r_*}=f(r)\frac{\rmd}{\rmd r}.
\end{align}
Assuming harmonic time dependence,
\begin{align}
 Q(t,r)=\psi_{l}(r)e^{\rmi\omega t},
\end{align}
the odd-parity perturbation equation takes the Schr\"odinger-type form
\begin{align}
 \left[ -\frac{\rmd^2}{\rmd r_*^2}+V^{\rm RW}_{l}(r) \right]\psi_{l}(r)=\omega^2\psi_{l}(r),
 \label{eq:RW_review_inserted}
\end{align}
with the Regge--Wheeler potential~\cite{Regge:1957td}
\begin{align}
 \vsup{V}{RW}_{l}(r)
 =f(r)
 \left(
 \frac{l(l+1)}{r^2}-\frac{6M}{r^3}
 \right).
 \label{eq:RWpotential}
\end{align}
This is the basic radial equation analyzed in the present work.

\subsection{Parity sectors and related master equations}

For later reference, it is useful to recall that Schwarzschild perturbations admit both odd- and even-parity master equations.
The odd-parity potential may be written more generally as~\cite{Kokkotas:1999bd}
\begin{align}
 \vsup{V}{odd}_{l}(r)
 =f(r)
 \left( \frac{l(l+1)}{r^2}+\frac{2\sigma M}{r^3} \right),
 \qquad
 \sigma=1-s^2,
\end{align}
where $s=0$, $1$, and $2$ correspond to scalar, electromagnetic, and gravitational perturbations, respectively.
For gravitational odd-parity perturbations $s=2$, this reduces to Eq.~\eqref{eq:RW_review_inserted}.
The even-parity (Zerilli) sector obeys a different $1$-dimensional wave equation with potential~\cite{Zerilli:1970wzz}
\begin{align}
 \vsup{V}{even}_{l}(r)
 =f(r)
 \frac{ 2n^2(n+1)r^3+6n^2Mr^2+18nM^2r+18M^3 }{ r^3(nr+3M)^2 },
 \qquad
 n=\frac{(l-1)(l+2)}{2}.
\end{align}
Although the present paper focuses on the odd-parity Regge--Wheeler problem, these related potentials are useful when discussing extensions of the complex-scaling framework beyond the simplest benchmark.

\subsection{Quasinormal-mode boundary conditions and Leaver benchmark}

Quasinormal modes (QNMs) are defined by purely ingoing behavior at the event horizon and purely outgoing behavior at spatial infinity.
Since $\vsup{V}{RW}_{l}(r)\to 0$ as $r_*\to\pm\infty$, the asymptotic
solutions behave as
\begin{align}
 \psi_{l}(r)\sim e^{-\rmi\omega r_*}
 \qquad r_*\to-\infty,
\end{align}
and
\begin{align}
 \psi_{l}(r)\sim e^{+\rmi\omega r_*}
 \qquad r_*\to+\infty.
\end{align}
In the Schwarzschild problem these conditions can be implemented very accurately by Leaver's continued-fraction method, which therefore provides an important benchmark for the complex-scaling calculations reported below.
For convenience, numerical values used for comparison are collected in Appendix~\ref{sec:leaver}.

\subsection{Extremal Reissner--Nordstr\"om black hole}
The Reissner--Nordstr\"om black hole is a charged generalization of the Schwarzschild solution with a charge $Q$.
The metric is modified by replacing $f(r)$ in Eq.~\eqref{eq:Schwarzschild-metric} with
\begin{align}
    \vsub{f}{RN}(r)=1-\frac{2M}{r}+\frac{Q^2}{r^2},
\end{align}
while the angular part of the metric is unchanged.
The charge $Q$ has a maximum value, $|Q|\leq M$. If maximally charged such that $|Q|=M$ such a Reissner--Nordstr\"om black hole is called extremal.

Gravitational perturbations with the Reissner--Nordstr\"om background have been studied in Refs.~\cite{Zerilli:1974ai,Moncrief:1974gw, Moncrief:1974ng}.
In particular, its odd-parity perturbation obeys Eq.~\eqref{eq:RW_review_inserted} but the potential differs from the Regge--Wheeler potential.
In the Reissner--Nordstr\"om case, the relation between $r_*$ and $r$ is obtained by solving
\begin{align}
    \frac{dr}{dr_*} = \frac{\Delta}{r^2},
\end{align}
and the corresponding potential is given by~\cite{Chandrasekhar:book}
\begin{align}
    \vsup{V}{odd}_{\mathrm{RN},s}(r)=\frac{\Delta}{r^5}\left(A r-q_{s}+\frac{p_sQ^2}{r}\right),
\end{align}
where
\begin{align}
    \Delta&=r^2-2Mr+Q^2, \\
    A&=l(l+1),\\
    q_0&=2,\\
    q_1&=3M-\sqrt{9M^2+4Q^2(l-1)(l+2)},\\
    q_2&=3M+\sqrt{9M^2+4Q^2(l-1)(l+2)}.\\
    p_0&=2,\\
    p_1&=p_2=4.
\end{align}
This potential provides the Zerilli--Moncrief equation.\footnote{%
Related charged backgrounds have also been considered in asymptotically AdS
settings, for example in studies of Reissner--Nordstr\"om--AdS black holes
\cite{Berti:2003ud,Liu:2025reu}.}
$\vsup{V}{odd}_{\mathrm{RN},0}$, $\vsup{V}{odd}_{\mathrm{RN},1}$, and $\vsup{V}{odd}_{\mathrm{RN},2}$ denote the potentials contributing to scalar, electromagnetic, and gravitational perturbations, respectively.\footnote{Note that the Reissner--Nordstr\"om metric is a solution to Einstein--Maxwell equations. The electromagnetic perturbations must be considered as well as the gravitational perturbations.}
In the limit $Q\to0$, $\vsup{V}{odd}_{2}$ leads to the Regge--Wheeler potential~\eqref{eq:RWpotential}.

Finally, let us comment on the applicability of Leaver's method to the Reissner--Nordstr\"om case, since this point motivates our CSM-based approach.
For a nonextremal Reissner--Nordstr\"om black hole, the Zerilli--Moncrief equation still admits the standard QNM boundary-value problem, and one may formulate a Leaver-type continued-fraction method.
In that case, the geometry has two distinct horizons,
\begin{align}
    r_{\pm}=M\pm\sqrt{M^2-Q^2},
\end{align}
so that the radial equation has regular singular points at $r=r_+$ and $r=r_-$, and the solution can be expanded in a Frobenius series around the outer horizon (see Appendix~\ref{sec:leaver}).

This situation changes qualitatively in the extremal limit $|Q|=M$, where the two horizons merge.
Because of this degeneration of the horizon structure, the nature of the singularity at the horizon is altered, and the radial equation is no longer in the standard form assumed in the naive Leaver construction.
In particular, the degenerate horizon gives rise to an irregular singular point, so that the usual Frobenius expansion about the outer horizon is not available in its standard form.
For this reason, the original continued-fraction method cannot be applied naively to extremal Reissner--Nordstr\"om black holes, and one must either substantially reformulate the method or adopt a different framework.
This is one of the reasons why it is worthwhile to develop a formulation, such as CSM, that is less directly tied to the Frobenius structure of the radial equation.

\section{Direct-measurement approach to resonance poles in scattering theory}\label{sec:csm}
As in quantum mechanics, many problems of interest are governed by Schr\"odinger-type equations, that is, elliptic-regular differential equations.
The Regge--Wheeler or Zerilli--Moncrief equation is also one of Schr\"odinger-type wave equations.
In this section, let us review a well-established technique, complex scaling method (CSM), to solve a general (time-independent) Schr\"odinger equation and to analyze quantum resonance/scattering theory.

\subsection{Theoretical and mathematical overview of CSM}

Since black-hole QNMs are resonances rather than ordinary bound states, their mode functions satisfy outgoing boundary conditions and are not square-integrable in the usual sense.
For a Schr\"odinger-type radial equation,\footnote{%
In the standard formulation, the radial function is often written as~$\psi(r)$.
In the present work, however, the complex scaling is implemented in the tortoise coordinate $r_*$, so we rewrite the wave function as~$\psi(r_*)$ in the theoretical discussion for simplicity.
In the numerical implementation, the potential is still evaluated as~$V(r(r_*))$; for the Schwarzschild case, the inverse relation $r=r(r_*)$ is written explicitly in terms of the Lambert $W$ function.}
\begin{align}
    \left[ -\frac{\rmd^2}{\rmd r_*^2}+V(r) \right]\psi(r_*)
    = \omega^2 \psi(r_*),
    \label{eq:csm-review-rw-like}
\end{align}
the QNM boundary conditions require
\begin{align}
    \psi(r_*) \sim
    \begin{cases}
        e^{-\rmi\omega r_*} & r_* \to -\infty,\\
        e^{+\rmi\omega r_*} & r_* \to +\infty,
    \end{cases}
\end{align}
with $\im \omega < 0$ for decaying modes.
Because of the sign of $\im \omega$, these solutions typically grow exponentially along at least one asymptotic direction and hence do not belong to the standard Hilbert space $L^2$.
This is the basic reason why QNMs are naturally interpreted as resonance poles of the analytically continued resolvent or S-matrix, rather than as ordinary normal modes \cite{Siegert:1939,Newton:1960}.

The complex scaling method (CSM) provides a way to convert this resonance problem into a non-Hermitian spectral problem~\cite{Myo:2014ypa}.
In the simplest global version, one performs the complex dilation
\begin{align}
    r_* \mapsto r_* e^{\rmi\theta},
    \qquad
    0 < \theta < \frac{\pi}{2},
\end{align}
which induces a non-unitary similarity transformation
\begin{align}
    H_\theta = U_\theta H U_\theta^{-1}.
\end{align}
For a $1$-dimensional Schr\"odinger operator $H=-\rmd^2/\rmd r_*^2+V(r)$, the transformed operator takes the schematic form
\begin{align}
    H_\theta
    =
    - e^{-2\rmi\theta}\frac{\rmd^2}{\rmd r_*^2}
    + V(r(r_* e^{\rmi\theta})),
    \label{eq:csm-review-global}
\end{align}
provided the potential admits a suitable analytic continuation.
$r=r(r_*)$ is understood through the inverse tortoise-coordinate relation.
In a nutshell, since the QNM boundary condition is changed by, for instance,
\begin{align}
    e^{\pm\rmi\omega r_*}
    \mapsto e^{\pm\rmi(\re \omega+\rmi\im \omega) r_* e^{\rmi\theta}}
    = e^{\mp (\re \omega \sin\theta + \im \omega \cos\theta) r_*}
    e^{\pm \rmi (\re \omega \cos\theta - \im \omega \sin\theta) r_*},
\end{align}
the complex-scaled wave function converges if $\re \omega \sin\theta + \im \omega \cos\theta > 0$, or $\theta > \tan^{-1} (-\im \omega/\re \omega)$.
\revone{The exposure condition is a necessary condition for a resonance to appear as an $L^2$ eigenstate at a given scaling angle. Consequently, the absence of an isolated eigenvalue at a smaller value of $\theta$ does not imply the absence of the corresponding QNM. In a reference-free calculation, one must scan the scaling angle over the analytically and numerically admissible interval and track stationary isolated branches. The present Gaussian representation further imposes $\theta<\pi/4$, and therefore cannot expose resonances whose required exposure angle exceeds this value. We accordingly make no claim of global completeness of the QNM spectrum.}
The mathematical foundation is the Aguilar--Balslev--Combes (ABC) theorem \cite{Aguilar:1971ve,Balslev:1971vb}, together with subsequent refinements~\cite{Simon1972BalslevCombes}.
Its essential consequences are the following.

First, resonance states satisfying outgoing boundary conditions become square-integrable after complex scaling.
Second, bound-state and resonance eigenvalues are independent of $\theta$ within the allowed analyticity wedge.
Third, the continuous spectrum is rotated into the lower half of the complex energy plane.
For the eigenvalue problem \eqref{eq:csm-review-rw-like}, the natural spectral parameter is
\begin{align}
    \lambda = \omega^2,
\end{align}
and the continuum on the positive real $\lambda$ axis is rotated by an angle
$2\theta$.
In this way, resonances that were hidden on the unphysical sheet become exposed
as isolated complex eigenvalues of the non-Hermitian operator $H_\theta$
\cite{Moiseyev:1998gjp,Myo:2014ypa}.\footnote{%
For black-hole perturbations, however, a global scaling of the entire radial coordinate may be unnecessarily strong.
What is actually problematic is the asymptotic region, where the outgoing-wave behavior diverges.
This motivates the use of exterior complex scaling (ECS), in which the coordinate is left unchanged in an inner region and rotated only beyond a finite onset radius $R_0$ \cite{Simon1979ECS}.
A standard choice is
\begin{align}
    g_{\theta,R_0}(r_*)
    =
    \begin{cases}
        r_*, & r_* \le R_0,\\
        R_0 + (r_*-R_0)e^{\rmi\theta}, & r_* > R_0,
    \end{cases}
    \label{eq:ecs-map-review}
\end{align}
or the analogous deformation on both asymptotic sides when needed.
The corresponding ECS operator is then obtained by replacing $r_*$ with $g_{\theta,R_0}(r_*)$ and transforming derivatives accordingly.
The advantage of ECS is that it preserves the interaction region while regularizing only the asymptotic outgoing sector.
This is especially natural in black-hole problems, where one wishes to keep the near-potential-barrier region intact and modify only the asymptotic domain responsible for the non-normalizable QNM behavior.
}

Operationally, the role of complex scaling is therefore not merely to provide a numerical trick.
Rather, it changes the functional setting of the problem so that resonances can be treated on the same footing as discrete eigenstates of a non-Hermitian
operator.
This point is particularly important when one wants to discuss not only individual QNM frequencies but also the relation between resonant and continuum sectors in a common spectral language.
In that setting, one works with right and left eigenfunctions,
\begin{align}
    H_\theta \ket{\psi_n^R}
    &= \lambda_n \ket{\psi_n^R},
    \\
    H_\theta^\dagger \ket{\psi_n^L}
    &= \lambda_n^* \ket{\psi_n^L},
\end{align}
which satisfy a biorthogonal normalization,
\begin{align}
    \braket{\psi_m^L | \psi_n^R}
    = \delta_{mn}.
\end{align}
The resolvent then admits the formal spectral representation
\begin{align}
    (z-H_\theta)^{-1}
    \sim
    \sum_n
    \frac{ \ket{\psi_n^R}\bra{\psi_n^L} }{ z-\lambda_n }
    + \text{continuum contribution},
    \label{eq:csm-review-resolvent}
\end{align}
which makes explicit how isolated resonances and the rotated continuum enter a single non-Hermitian spectral decomposition~\cite{Moiseyev:1998gjp,Myo:2014ypa}.

From the viewpoint of the present paper, this framework is attractive for two
reasons.
First, it gives a direct eigenvalue formulation of black-hole QNMs without imposing the outgoing boundary condition by hand at infinity in the original real coordinate.
Second, it is less tightly tied to the specific Frobenius or continued-fraction structure of a particular radial equation.
This flexibility is important when one moves from the Schwarzschild Regge--Wheeler problem to the Reissner--Nordstr\"om family, and especially
toward the extremal limit, where the analytic structure at the horizon becomes more delicate and the naive Leaver construction is no longer available in its usual form.
For this reason, CSM and ECS provide a natural framework for the present analysis of QNM spectra in both Schwarzschild and Reissner--Nordstr\"om black holes.

In addition to isolated resonance poles, the same framework should in principle also allow one to study continuum observables, such as the continuum level density, which may be relevant to branch-cut contributions
and late-time tails (see Appendix~\ref{app:cld}).

\subsection{Numerical implementation and some schemes of choosing bases}
\subsubsection{Variational procedure to eigenvalue problem}
We now turn to the practical implementation of the complex-scaled eigenvalue problem.
The purpose of this section is not to introduce a new mathematical structure, but rather to explain how the complex-scaled radial equation is converted into a finite-dimensional matrix problem suitable for numerical computation.

As discussed in the previous subsection, the essential effect of complex scaling is to transform the outgoing-wave resonance problem into a non-Hermitian eigenvalue problem.
For the Schwarzschild case, this begins with the Regge--Wheeler equation,
which has the Schr\"odinger-type form
\begin{align}
    \left[ -\frac{\rmd^2}{\rmd r_*^2} + V_l(r) \right]
    \psi_l(r_*)
    =
    \omega^2 \psi_l(r_*).
    \label{eq:rw_schr}
\end{align}
After complex scaling, the kinetic term acquires the factor $e^{-2 \rmi \theta}$, while the potential is evaluated along the deformed contour.
The resulting equation can then be treated as an eigenvalue problem for the complex-scaled operator.

To solve this problem numerically, we expand the complex-scaled wave function in a finite set of square-integrable basis functions and apply the Rayleigh--Ritz variational procedure.
In the present work, we adopt several types of Gaussian basis functions of both even and odd parity.
This choice is convenient because such a basis is analytically simple, provides stable matrix elements for the kinetic term, and allows a flexible coverage of the tortoise-coordinate domain through a geometric progression of Gaussian ranges.
The complex-scaled differential equation is thereby reduced to a generalized matrix eigenvalue problem for the expansion coefficients.

This basis-expansion strategy is particularly suitable for the present purpose.
First, it is fully compatible with the CSM framework, in which resonance wave functions become effectively $L^2$-integrable after complex scaling.
Second, it does not rely on a Frobenius expansion tied to a specific horizon structure, unlike continued-fraction approaches of Leaver type.
This point is important for our broader goal of treating not only the Schwarzschild problem but also the Reissner--Nordstr\"om family within a common framework.

In what follows, we first construct the matrix representation of the complex-scaled Regge--Wheeler operator in a simple Gaussian basis.
We then explain how the generalized complex eigenvalue problem is solved in practice and how physically relevant QNM frequencies are identified among the resulting eigenvalues.

\subsubsection{Demonstration in a simpler case: $e^{-\alpha r_*^2}$ and $r_* e^{-\alpha r_*^2}$}

As a first implementation of the complex scaling method, we discretize the
complex-scaled Regge--Wheeler equation in a Gaussian basis.
The purpose of this subsection is mainly illustrative:
it provides a direct realization of the CSM eigenvalue problem in terms of
explicit basis functions and matrix elements, and it serves as a useful
reference point before introducing more efficient basis choices in the next
subsubsection.

We begin with the Regge--Wheeler equation~\eqref{eq:rw_schr} with~$\vsup{V}{RW}_l(r)$~\eqref{eq:RWpotential}.
Under complex scaling, the equation is transformed into
\begin{align}
    \left[
        -e^{-2\rmi\theta}\frac{\rmd^2}{\rmd r_*^2}
        + V_l^\theta(r_*)
    \right]
    \psi_l^\theta(r_*e^{\rmi\theta})
    =
    (\omega^\theta)^2 \psi_l^\theta(r_*e^{\rmi\theta}),
    \label{eq:rw-csm-gaussian}
\end{align}
where $V_l^\theta$ denotes the potential evaluated along the
complex-rotated contour.

To solve Eq.~\eqref{eq:rw-csm-gaussian}, we use the Gaussian expansion method~\cite{Hiyama:2003cu} and expand the complex-scaled
wave function in square-integrable Gaussian basis functions.
Since, in practice, an expansion restricted to even functions alone is not
sufficient to reproduce the quasinormal frequencies accurately, we include
both even- and odd-parity Gaussian functions:
\begin{align}
    \psi_l^\theta(r_*e^{\rmi\theta})
    =
    \sum_{i=1}^{i_{\max}}
    C_{l i}^{\theta,\mathrm{(even)}}
    \phi_i^{\mathrm{(even)}}(r_*)
    +
    \sum_{i=1}^{i_{\max}}
    C_{l i}^{\theta,\mathrm{(odd)}}
    \phi_i^{\mathrm{(odd)}}(r_*).
    \label{eq:gaussian-expansion}
\end{align}
$i_{\max}$ is the truncation order of a basis.
We choose
\begin{align}
    \vsup{\phi}{(even)}_i(r_*)
    &= \vsup{N}{(even)}_i
    \exp\!\left[ -\left(\frac{r_*}{r_i}\right)^2 \right]
    = \vsup{N}{(even)}_i e^{-\alpha_i r_*^2},
    \\
    \vsup{\phi}{(odd)}_i(r_*)
    &= \vsup{N}{(odd)}_i
    \left(\frac{r_*}{r_i}\right) \exp\!\left[ -\left(\frac{r_*}{r_i}\right)^2 \right]
    = \vsup{N}{(odd)}_i (\sqrt{\alpha_i}\, r_*) e^{-\alpha_i r_*^2},
\end{align}
with
\begin{align}
    \alpha_i = \frac{1}{r_i^2}.
\end{align}
The normalization constants are chosen as
\begin{align}
    \vsup{N}{(even)}_i
    = \left( \frac{2\alpha_i}{\pi} \right)^{1/4},
    \qquad
    \vsup{N}{(odd)}_i
    = \left( \frac{2^{5}\alpha_i}{\pi} \right)^{1/4}.
\end{align}
The Gaussian ranges are taken in geometric progression,
\begin{align}
    r_i = r_{*0}\, a^{\,i-1},
    \qquad
    a = \left( \frac{r_{*\max}}{r_{*0}} \right)^{1/(i_{\max}-1)},
    \label{eq:gaussian-ranges}
\end{align}
so that the basis efficiently covers both short- and long-distance regions in the tortoise coordinate.

Applying the Rayleigh--Ritz variational procedure to the expansion
\eqref{eq:gaussian-expansion}, we obtain a generalized complex eigenvalue
problem,
\begin{align}
    \sum_j \left[ H_{ij}^\theta - (\omega^\theta)^2 N_{ij} \right]
    C_j^\theta
    = 0,
    \label{eq:generalized-evp-gaussian}
\end{align}
where the Hamiltonian matrix has the block form
\begin{align}
    H_{ij}^\theta
    =
    \begin{pmatrix}
        T_{ij}^{\theta,\mathrm{(even)}}
        +
        V_{l,ij}^{\theta,\mathrm{(even)}}
        &
        V_{l,ij}^{\theta,\mathrm{(even\mbox{-}odd)}}
        \\
        V_{l,ij}^{\theta,\mathrm{(odd\mbox{-}even)}}
        &
        T_{ij}^{\theta,\mathrm{(odd)}}
        +
        V_{l,ij}^{\theta,\mathrm{(odd)}}
    \end{pmatrix},
\end{align}
and the overlap matrix is
\begin{align}
    N =
    \begin{pmatrix}
        \vsup{N}{(even)}_{ij} & 0 \\
        0 & \vsup{N}{(odd)}_{ij}
    \end{pmatrix}.
\end{align}

The overlap matrix elements are given analytically by
\begin{align}
    \vsup{N}{(even)}_{ij}
    &= \int_{-\infty}^{\infty} dr_*\,
    \vsup{\phi}{(even)}_i(r_*) \vsup{\phi}{(even)}_j(r_*)
    = \vsup{N}{(even)}_i \vsup{N}{(even)}_j
    \frac{\sqrt{\pi}}{(\alpha_i+\alpha_j)^{1/2}},
    \\
    N_{ij}^{\mathrm{(odd)}}
    &= \int_{-\infty}^{\infty} dr_*\,
    \vsup{\phi}{(odd)}_i(r_*) \vsup{\phi}{(odd)}_j(r_*)
    = \vsup{N}{(odd)}_i \vsup{N}{(odd)}_j \sqrt{\alpha_i\alpha_j}
    \frac{\sqrt{\pi}}{2(\alpha_i+\alpha_j)^{3/2}}.
\end{align}
Similarly, the kinetic-energy matrix elements can also be evaluated in closed
form:
\begin{align}
    T_{ij}^{\theta,\mathrm{(even)}}
    &= e^{-2\rmi\theta} \vsup{N}{(even)}_i \vsup{N}{(even)}_j
    \frac{ 2\sqrt{\pi}\,\alpha_i\alpha_j }{ (\alpha_i+\alpha_j)^{3/2} },
    \\
    T_{ij}^{\theta,\mathrm{(odd)}}
    &=
    e^{-2\rmi\theta} \vsup{N}{(odd)}_i \vsup{N}{(odd)}_j
    \frac{ 3\sqrt{\pi}\,(\alpha_i\alpha_j)^{3/2} }{ (\alpha_i+\alpha_j)^{5/2} }.
\end{align}

The potential matrix elements are expressed as\footnote{%
In the present implementation, the practical upper bound on the scaling angle is not set by the resonance physics itself, but by the convergence of the basis representation.
The Gaussian basis, $\exp[- (\alpha_i+\alpha_j) e^{-2 \rmi \theta} r_*^2]$, converges only if $\cos2\theta>0$, i.e., $\theta<\pi/4$.
More generally, in black-hole problems the inverse relation $r=r(r_*)$ may develop nontrivial branch structures after analytic continuation, so that the allowed range of the scaling angle can be restricted not only by the basis representation but also by the branch structure of $r(r_*)$ itself.}
\begin{align}
    V_{l,ij}^{\theta,\mathrm{(even)}}
    &= e^{-\rmi\theta} \vsup{N}{(even)}_i \vsup{N}{(even)}_j
    \int_{-\infty}^{\infty} dr_*\,
    V_l(r)\, e^{-(\alpha_i+\alpha_j)e^{-2\rmi\theta}r_*^2},
    \\
    V_{l,ij}^{\theta,\mathrm{(odd)}}
    &= e^{-3\rmi\theta} \vsup{N}{(odd)}_i \vsup{N}{(odd)}_j
    \sqrt{\alpha_i\alpha_j}\int_{-\infty}^{\infty} dr_*\,
    V_l(r)\, r_*^2 e^{-(\alpha_i+\alpha_j)e^{-2\rmi\theta}r_*^2},
    \\
    V_{l,ij}^{\theta,\mathrm{(even\mbox{-}odd)}}
    &= e^{-2\rmi\theta} \vsup{N}{(even)}_i \vsup{N}{(odd)}_j
    \sqrt{\alpha_j}\int_{-\infty}^{\infty} dr_*\,
    V_l(r)\, r_* e^{-(\alpha_i+\alpha_j)e^{-2\rmi\theta}r_*^2},
    \\
    V_{l,ij}^{\theta,\mathrm{(odd\mbox{-}even)}}
    &= V_{l,ji}^{\theta,\mathrm{(even\mbox{-}odd)}}.
\end{align}
Here $r=r(r_*)$ is understood through the inverse tortoise-coordinate relation.
For Schwarzschild spacetime, this relation may be written explicitly in terms
of the Lambert $W$ function, which is convenient for the numerical evaluation
of the potential matrix elements.

The resulting matrix is complex symmetric, and the quasinormal frequencies are
obtained by solving the generalized complex eigenvalue problem
\eqref{eq:generalized-evp-gaussian}.
Although this Gaussian discretization is not yet the most efficient basis for
the present problem, it is conceptually transparent and provides a direct
illustration of how the CSM formulation can be implemented in practice.
This representation therefore serves as a useful starting point before turning
to improved basis choices in the next subsubsection.

\subsubsection{Variations of Gaussian basis functions and diagonalization}

In addition to the Gaussian basis introduced in the previous subsubsection, it is
useful to consider several alternative basis sets.
The purpose of this subsubsection is to summarize the candidate basis functions
used in our exploration of the CSM eigenvalue problem.
While the real range Gaussian basis already provides a direct implementation of
the CSM eigenvalue problem, other basis choices can offer improved flexibility
for oscillatory or resonance-like wave functions.
\begin{itemize}
    \item \textbf{Real range Gaussian}
    
    The wave function is expanded as
    \begin{align}
        \psi^\theta(r_* e^{\rmi\theta})
        &=
        \sum_{i=1}^{i_{\max}}
        \vsup{C}{even}_i \vsup{N}{even}_i e^{-\alpha_i r_*^2}
        +
        \sum_{i=1}^{i_{\max}}
        \vsup{C}{odd}_i \vsup{N}{odd}_i (\sqrt{\alpha_i}\, r_*) e^{-\alpha_i r_*^2}.
    \end{align}

    

    \item \textbf{Polynomial $\times$ real range Gaussian}
    
    The wave function is expanded as
    \begin{align}
        \psi^\theta(r_* e^{\rmi\theta})
        &=
        \sum_{p=0}^{p_{\max}}
        \sum_{i=1}^{i_{\max}}
        C_i^{p} N_i^{p}
        \left( \sqrt{\alpha_i}\, r_* \right)^p
        e^{-\alpha_i r_*^2}.
    \end{align}
    $p_{\max}$ is the highest degree of the polynomial of~$r_*$;
    for each~$i$, the basis is spanned by~$1$, $r_*$, $r_*^2$, \dots, $r_*^{p_{\max}}$ with the Gaussian factor.

    \item \textbf{sin/cos $\times$ real range Gaussian}
    
    The wave function is expanded as
    \begin{align}
        \psi^\theta(r_* e^{\rmi\theta})
        &=
        \sum_{i=1}^{i_{\max}}
        C_i^{\sin} N_i^{\sin}
        \sin\!\left( \sqrt{\alpha_i}\, r_* \right)
        e^{-\alpha_i r_*^2}
        \notag\\
        &\quad
        +
        \sum_{i=1}^{i_{\max}}
        C_i^{\cos} N_i^{\cos}
        \cos\!\left( \sqrt{\alpha_i}\, r_* \right)
        e^{-\alpha_i r_*^2}.
    \end{align}
\end{itemize}
The Real range Gaussian basis corresponds to the Polynomial $\times$ real range Gaussian one with $p_{\max}=1$.

Among these candidates, the Polynomial or sin/cos $\times$ real range Gaussian basis is expected to be
particularly effective for representing oscillatory behavior while maintaining
the analytic convenience of Gaussian matrix elements.

After choosing a basis set, we represent the complex-scaled wavefunction as
\begin{align}
    \psi^\theta(r_* e^{\rmi\theta})
    = \sum_i c_i^\theta \phi_i(r_*),
\end{align}
where $\{\phi_i\}$ denotes one of the basis sets introduced above.
Substituting this expansion into the complex-scaled wave equation and applying
the Rayleigh--Ritz procedure, we obtain the generalized eigenvalue problem
\begin{align}
    H_\theta c
    =
    \lambda N c,
    \qquad
    \lambda = \omega^2,
    \label{eq:gevp-diag}
\end{align}
where $H_\theta$ is the complex-scaled Hamiltonian matrix,
$N$ is the overlap matrix of the non-orthogonal basis,
and $c$ is the coefficient vector.

For basis sets with sufficiently mild overlap, one may solve
Eq.~\eqref{eq:gevp-diag} directly as a generalized complex eigenvalue problem.
In practice, however, the basis functions used here are often strongly
non-orthogonal, and the overlap matrix may contain nearly linearly dependent
directions.
To improve numerical stability, we therefore orthogonalize the basis before
the final diagonalization whenever necessary.

More precisely, we first diagonalize the overlap matrix,
\begin{align}
    N = U K U^\dagger,
\end{align}
where
\begin{align}
    K = \diag(\kappa_1,\kappa_2,\dots)
\end{align}
contains the eigenvalues of $N$, and $U$ is the corresponding eigenvector
matrix.
Small eigenvalues of $N$ indicate nearly redundant directions in the truncated
basis.
We therefore discard basis components whose overlap eigenvalues satisfy
\begin{align}
    |\kappa_i| \le \epsilon_N,
\end{align}
with $\epsilon_N$ a numerical cutoff.
We set $\epsilon_N=10^{-5}$, a value often adopted empirically.
After this truncation, the Hamiltonian is transformed to the orthonormalized
basis according to
\begin{align}
    \Tilde{H}_\theta
    =
    K^{-1/2} U^\dagger H_\theta U K^{-1/2},
    \label{eq:orth-ham}
\end{align}
where only the retained subspace is understood.
The generalized eigenvalue problem~\eqref{eq:gevp-diag} is thereby reduced to
the ordinary eigenvalue problem
\begin{align}
    \Tilde{H}_\theta \Tilde{c} = \lambda \Tilde{c}.
    \label{eq:ordinary-evp}
\end{align}

We then diagonalize the complex matrix
$\Tilde{H}_\theta$
and obtain a set of complex eigenvalues $\lambda_k$.
The corresponding quasinormal-frequency candidates are defined by
\begin{align}
    \omega_k = \sqrt{\lambda_k},
\end{align}
where the physically relevant branch is selected by the usual QNM convention,
namely
\begin{align}
    \im \omega_k < 0.
\end{align}

In the present implementation, this orthogonalization step is particularly
important for the polynomial and trigonometric Gaussian bases, for which the
overlap matrix can become ill-conditioned as the basis size increases.
The orthogonalized formulation substantially improves the numerical stability
of the diagonalization and suppresses spurious directions associated with
near-linear dependence in the original non-orthogonal basis.

It should be emphasized that the output of a single diagonalization is not yet
sufficient to identify the physical QNMs.
The resulting spectrum generally contains not only resonance candidates but
also discretized continuum states.
For this reason, the physical QNM frequencies are identified only after the
stability analysis described below, in particular through their dependence on
the scaling angle, basis size, and basis parameters.

\revone{For completeness, let us summarize the computational scaling of the present
finite-basis implementation.  For the Polynomial $\times$ real range
Gaussian basis, the raw basis dimension is
\begin{equation}
 \vsub{D}{raw}
 =
 (p_{\max}+1)i_{\max}.
\end{equation}
After the nearly linearly dependent directions of the overlap matrix have
been removed, we denote the retained dimension by
$\vsub{D}{ret}\leq\vsub{D}{raw}$.}

\revone{The overlap and Hamiltonian matrices require
$O(\vsub{D}{raw}^2)$ storage. The full dense eigendecomposition of the overlap matrix has an
asymptotic cost of $O(\vsub{D}{raw}^3)$. The dense basis
transformation costs at most
$O(\vsub{D}{raw}^2\vsub{D}{ret}+\vsub{D}{raw}\vsub{D}{ret}^2)$, while the subsequent full
eigendecomposition of the orthogonalized complex Hamiltonian has an
asymptotic cost of $O(\vsub{D}{ret}^3)$. The construction of the potential matrix involves
$O(\vsub{D}{raw}^2)$ matrix elements, with a prefactor that depends on
the numerical quadrature used for each element.}

\revone{In the main calculations, we use $p_{\max}=3$ and
$i_{\max}=30$--$50$, corresponding to raw dimensions
$\vsub{D}{raw}=120$--$200$.  At these modest dimensions, the runtime of
an individual calculation is sufficiently short that hardware-dependent
and software-initialization overheads are comparable to the numerical
kernel.  For this reason, a wall-clock timing table would not provide a
robust or reproducible performance benchmark.  We instead report the
matrix dimensions and asymptotic scaling.  At substantially larger basis
sizes, the dense eigensolvers, which are the cubic-scaling steps, are
expected to control the computational growth.}

\revone{Calculations at different scaling angles and with different basis
parameters are mutually independent.  The total cost of a parameter scan
therefore grows approximately linearly with the number of parameter sets,
and the scan can be parallelized directly.}

\subsection{\revone{Reference-free identification and stability analysis of QNM frequencies}}\label{sec:stability_qnm}




\revone{
The eigenvalues obtained from a single diagonalization generally contain
resonance candidates, discretized continuum states, and basis-dependent
artifacts. We therefore perform the mode-selection step using only the
complex-scaled spectra, before consulting any previously known QNM frequency.
The procedure described below is implemented as an automated scan over the
computed spectra.}

\revone{Let $\omega_a(\theta_j)$ be the frequencies obtained at the scaling angle
$\theta_j$, and define
\begin{equation}
  E_a(\theta_j)=\omega_a(\theta_j)^2.
\end{equation}
We first restrict the search to the physical QNM quadrant and to a finite
frequency window,
\begin{equation}
  \re\omega_a>0,
  \qquad
  \im\omega_a<0,
  \qquad
  |\omega_a|\leq\omega_{\max}.
\end{equation}
For each continuum threshold $E_{\mathrm{th},\alpha}$, the complex-scaled
continuum is represented by the half-line
\begin{equation}
  \mathcal{R}_{\theta,\alpha}
  =
  \left\{
    E_{\mathrm{th},\alpha}
    +t\rme^{-2\rmi\theta}
    \mathrel{\big|}
    t\geq0
  \right\}.
\end{equation}
A frequency is retained as a candidate only when its energy lies inside the
exposed resonance wedge for at least one threshold,
\begin{equation}
  -2\theta
  <
  \Arg\left(E_a-E_{\mathrm{th},\alpha}\right)
  <
  0,
\end{equation}
and when it is separated from all rotated continuum rays. We quantify the
latter condition by
\begin{equation}
  \vsub{d}{ray}(E_a;\theta)
  =
  \min_{\alpha}
  \min_{t\geq0}
  \left|
    E_a-E_{\mathrm{th},\alpha}
    -t\rme^{-2\rmi\theta}
  \right|,
\end{equation}
and require
\begin{equation}
  \vsub{d}{ray}(E_a;\theta)
  >
  \vsub{\delta}{ray}.
\end{equation}}

\revone{The candidates obtained at different scaling angles are then clustered in the
complex $\omega$ plane. Starting from a reference angle, we associate a
candidate with the nearest candidate in each of the other spectra, provided
that
\begin{equation}
  \left|
    \omega_a(\theta_j)
    -
    \omega_k(\vsub{\theta}{ref})
  \right|
  <
  \delta_{\omega}.
\end{equation}
A cluster is retained only when it contains candidates from at least
$\vsub{N}{hit}$ sampled scaling angles. In the calculations reported below,
we use
\begin{equation}
  \omega_{\max}=5,
  \qquad
  \vsub{\delta}{ray}=10^{-3},
  \qquad
  \delta_{\omega}=10^{-2},
  \qquad
  \vsub{N}{hit}=2,
\end{equation}
in the dimensionless units used in the numerical calculation. The numerical values above are empirical working tolerances
used in the automated post-processing. They are not universal
thresholds for QNM identification.}

\revone{For a retained cluster containing $N_k$ frequencies, we define its mean and
componentwise spreads by
\begin{align}
  \bar{\omega}_k
  &=
  \frac{1}{N_k}
  \sum_{j=1}^{N_k}
  \omega_{k,j},
  \\
  \Delta_{k}^{(R)}
  &=
  \max_j
  \left|
    \re\omega_{k,j}
    -
    \re\bar{\omega}_k
  \right|,
  \\
  \Delta_{k}^{(I)}
  &=
  \max_j
  \left|
    \im\omega_{k,j}
    -
    \im\bar{\omega}_k
  \right|.
\end{align}
These spreads quantify the residual variation within the sampled cluster.
The same automated procedure is applied to spectra obtained with
different basis sizes, Gaussian ranges, and integration ranges whenever
the corresponding data are available. Agreement under these finite
parameter variations is a stability diagnostic and is not interpreted as proof
of asymptotic convergence.}

\revone{We use the following reporting convention. A mode is \textit{robustly
identified} when a persistent cluster is selected by the above procedure and
remains compatible under the tested basis and range variations. A mode is
reported as a \textit{tentative candidate} when a persistent cluster is found
but its parameter-variation spread is too large to support the intended
numerical precision. The label \textit{not identified} is used when one or more spectra
contain a marginal feature suggestive of a candidate, but no cluster
satisfies the isolation and cross-angle persistence criteria strongly
enough for a frequency to be quoted. This label does not imply that
the physical QNM is absent or fundamentally unresolved. A dash in the summary
tables simply indicates that no CSM frequency is quoted for that entry in the
present scan.}

\revone{When Leaver, Onozawa, or other reference frequencies are available, they are
consulted only after the above CSM-only extraction has been completed. Their
role is to provide an a posteriori assessment of numerical accuracy, not to
select the candidate clusters.}

%
%

\section{Numerical simulation of QNM via CSM}\label{sec:numerics}

\subsection{Benchmark strategy}


\revone{The Schwarzschild frequencies obtained from Leaver's continued-fraction
method, together with the other reference values summarized in
Appendix~\ref{sec:leaver}, are used exclusively for a posteriori benchmarking.
The mode-selection step is first performed according to the
reference-free procedure of Sec.~\ref{sec:stability_qnm}, using only the complex-scaled
spectra.}

\revone{After the CSM-only candidate extraction and reporting status have been
fixed, the reference frequencies are consulted to assess the numerical
accuracy of the CSM results. Agreement with a reference value is
therefore an a posteriori validation of the extracted candidate, rather
than an input to the mode-selection procedure.}

\subsection{Trial run of the methodology and comparison of basis dependence}

As a first trial run of the present methodology, we compare the complex-scaled spectra obtained with several basis sets.
The purpose of this comparison is twofold.
First, it provides a basic numerical check that the physically relevant QNM frequencies are reproduced in a reasonably stable manner across different basis choices.
Second, it illustrates that, although the detailed distribution of the nonphysical eigenvalues depends on the basis, the low-lying resonance candidates are much more robust.

We focus on the $l=2$ gravitational mode ($s=2$) for the Regge--Wheeler equation with $M=0.5$ and consider three representative basis sets: the Real range Gaussian basis (equivalently, the Polynomial $\times$ real range Gaussian basis with $p_{\max}=1$), the Polynomial $\times$ real range Gaussian basis with $p_{\max}=3$, and the sin/cos $\times$ real range Gaussian basis.
In each case, we diagonalize the complex-scaled Hamiltonian for nearby values of the scaling angle $\theta$ and compare the resulting eigenvalue distributions in the complex $\omega$ plane.
The stability of the lowest resonance candidates under such changes provides a basic diagnostic for identifying the physical QNMs.

\begin{figure}[ht]
    \centering
    \includegraphics[width=0.7\columnwidth]{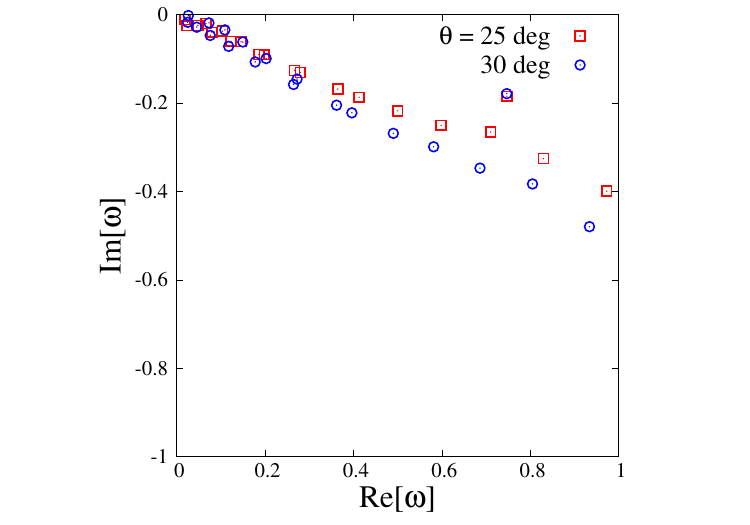}
    \includegraphics[width=0.7\columnwidth]{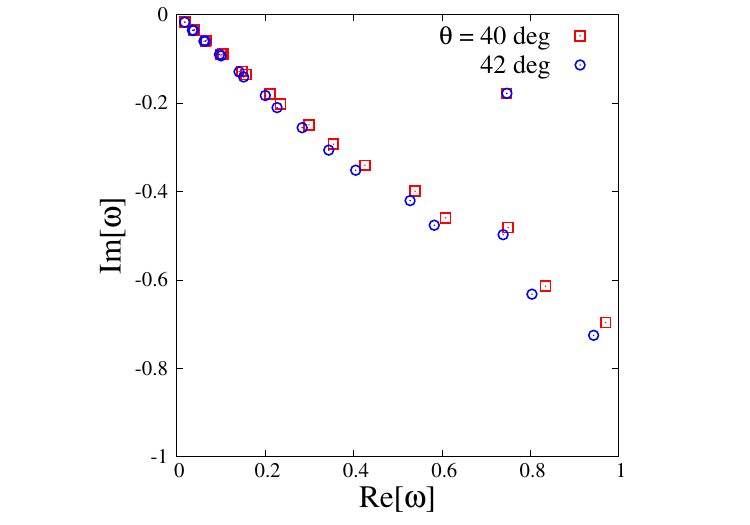}
    \caption{$l=2$, Real range Gaussian basis.
    The scaling angle is taken to be~$\theta=25^\circ$, $30^\circ$ on the upper panel, and $\theta=40^\circ$, $42^\circ$ on the lower panel.
    We take  $i_{\max}=30$, $r_{*0}=0.1$, and $r_{*\max}=60$.
    The sequence of points aligned approximately along the $\theta$-rotated direction represents the discretized continuum spectrum, while the isolated points on the real-axis side are interpreted as resonance eigenvalues, namely the QNM frequencies.
    The lowest resonance candidate is
    $\omega_1=0.747343-\rmi0.177958$ ($\theta=40^{\circ}$) and
    $\omega_1=0.747349-\rmi0.177940$ ($\theta=42^{\circ}$).
    The next candidate is
    $\omega_2=0.750229-\rmi0.481678$ ($\theta=40^{\circ}$) and
    $\omega_2=0.738955-\rmi0.497959$ ($\theta=42^{\circ}$).}
    \label{fig:ene_l2_poly_nmax1}
\end{figure}

Figure~\ref{fig:ene_l2_poly_nmax1} shows the result obtained with the real range Gaussian basis.
The scaling angle is taken to be $\theta=25^\circ$, $30^\circ$ in the upper panel, and $\theta=40^\circ$, $42^\circ$ in the lower panel.
As seen in the figure, a sequence of discrete eigenvalues appears roughly along a rotated branch in the complex $\omega$ plane.
This sequence represents the discretized continuum spectrum in the present finite-basis calculation.
By contrast, the isolated points on the real-axis side of this branch are interpreted as resonance eigenvalues, namely the QNM frequencies.
This is the standard spectral pattern expected in the CSM: continuum-related states move with the scaling angle and align along the rotated branch, whereas genuine resonance states remain relatively isolated from it and show much weaker $\theta$ dependence.
For this reason, the identification of QNM frequencies is based not on a single diagonalization result alone, but on whether an eigenvalue remains isolated from the rotated continuum and stable under changes of $\theta$.

The lowest mode in Fig.~\ref{fig:ene_l2_poly_nmax1} is already rather stable under the changes $\theta=25^\circ \to 30^\circ$ and $\theta=40^\circ \to 42^\circ$, while the second mode exhibits a visibly larger displacement for $\theta \gtrsim 40^\circ$.
This behavior is consistent with the general expectation that the fundamental QNM is more robust than higher-lying states against basis and scaling-angle variations.

\revone{The necessary exposure condition is given by
\begin{align}
    \theta > \arctan\left(\frac{-\im\omega}{\re\omega}\right) .
\end{align}
Applying this condition a posteriori to the Schwarzschild benchmark, the exposure angle of the second mode shown in Fig.~\ref{fig:ene_l2_poly_nmax1} is approximately $38.3^\circ$. Its absence at $\theta=25^\circ$ and $30^\circ$ is therefore expected and is not a numerical omission. By contrast, it becomes exposed at $\theta=40^\circ$ and $42^\circ$, although its weak localization near the exposure threshold makes it more sensitive to the finite basis.
Moreover, the Gaussian matrix elements require $\theta<\pi/4$. The next Schwarzschild overtone would require an exposure angle of
approximately $57.8^\circ$ and hence cannot be accessed within the
present Gaussian representation, irrespective of how densely the
interval below $45^\circ$ is scanned.}

\revone{We therefore do not claim global completeness of the QNM spectrum.
The present calculation probes only the spectral region exposed by
the sampled scaling angles and representable within the chosen
Gaussian basis.}

\revone{A mode whose required exposure angle lies outside the admissible
interval is described as \textit{not accessible within the present
implementation}. For a mode inside the accessible region, failure to
obtain a persistent candidate cluster is reported as \textit{not
identified in the present scan}. The latter wording is deliberately
weaker: a marginal feature may be suggestive of a candidate without
satisfying the quantitative isolation and persistence filters, and
its failure to be selected does not establish the absence of the
physical QNM.}

\revone{In practice, spectra generated for the selected scaling angles are
processed in a single automated batch using the selection and
clustering procedure of Sec.~3.3. The same procedure is applied to
additional basis sizes, Gaussian ranges, and integration ranges
whenever the corresponding spectra are available. Since the
individual diagonalizations are independent, the total cost grows
approximately linearly with the number of parameter sets, and the
scan can be parallelized directly.}

\begin{figure}[ht]
    \centering
    \includegraphics[width=0.7\columnwidth]{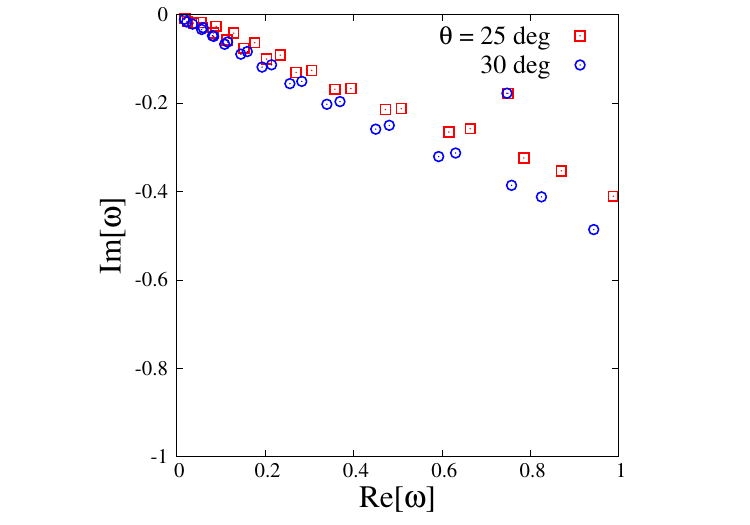}
    \includegraphics[width=0.7\columnwidth]{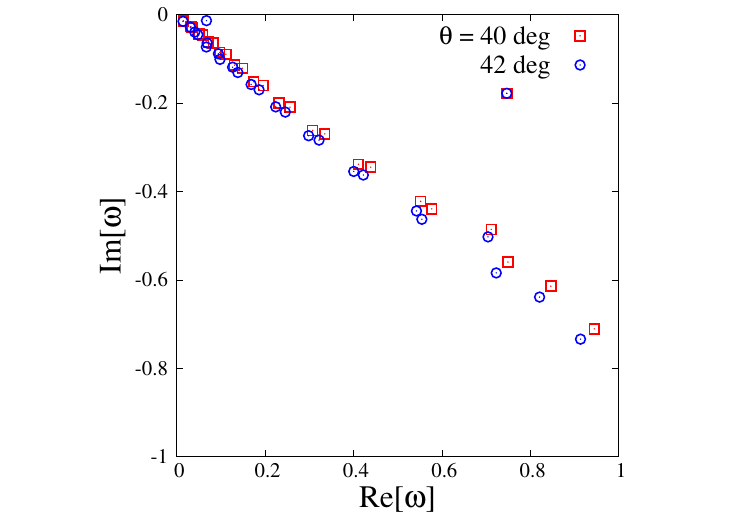}
    \caption{$l=2$, Polynomial $\times$ real range Gaussian basis with~$p_{\max}=3$.
    We take $i_{\max}=30$, $r_{*0}=0.1$, and $r_{*\max}=60$.
    The lowest resonance candidate is
    $\omega_1=0.747348-\rmi0.177926$ ($\theta=40^{\circ}$) and
    $\omega_1=0.747345-\rmi0.177925$ ($\theta=42^{\circ}$).
    The next candidate is
    $\omega_2=0.712209-\rmi0.485971$ ($\theta=40^{\circ}$),
    $\omega_2=0.704895-\rmi0.502870$ ($\theta=42^{\circ}$).}
    \label{fig:ene_l2_poly_nmax3}
\end{figure}

Figure~\ref{fig:ene_l2_poly_nmax3} shows the corresponding result for the Polynomial $\times$ real range Gaussian basis with $p_{\max}=3$.
Compared with the previous case, the fundamental mode is again highly stable, whereas the second mode still shows a more noticeable $\theta$ dependence.
At the same time, the polynomial basis yields a visibly better reproduction of the low-lying frequencies than the simple real range Gaussian basis, in particular for the first overtone.
This indicates that the present CSM formulation robustly captures the lowest QNM, while higher modes remain more sensitive to the details of the truncated basis representation.

\begin{figure}[ht]
    \centering
    \includegraphics[width=0.7\columnwidth]{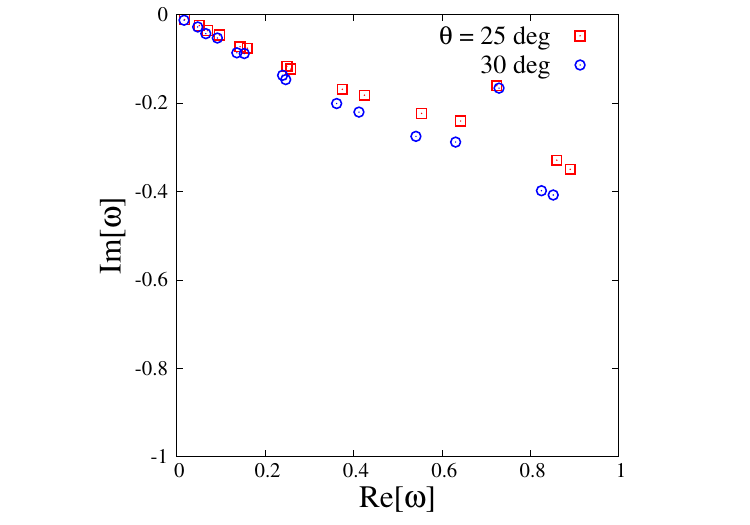}
    \includegraphics[width=0.7\columnwidth]{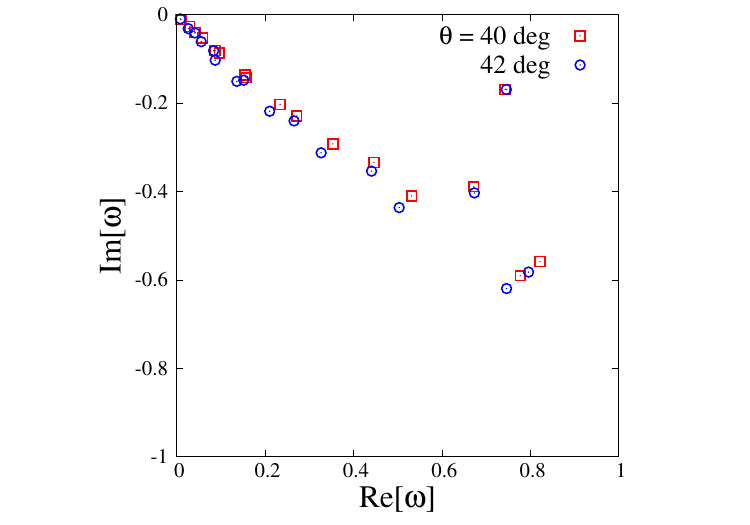}
    \caption{$l=2$, sin/cos $\times$ real range Gaussian basis.
    We take $i_{\max}=30$, $r_{*0}=0.1$, and $r_{*\max}=80$.
    The lowest resonance candidate is
    $\omega_1=0.743544-\rmi0.168990$ ($\theta=40^\circ$) and
    $\omega_1=0.746374-\rmi0.169193$ ($\theta=42^\circ$).
    The second-mode candidate is
    $\omega_2=0.672161-\rmi0.389585$ ($\theta=40^\circ$) and
    $\omega_2=0.673930-\rmi0.403464$ ($\theta=42^\circ$).}
    \label{fig:ene_l2_trigo}
\end{figure}

Finally, Fig.~\ref{fig:ene_l2_trigo} displays the result for the sin/cos $\times$ real range Gaussian basis.
This basis is designed to incorporate more explicit oscillatory structure into the trial space and is therefore expected to be advantageous for resonance wave functions.
However, in practice its performance is more delicate.
Although the detailed eigenvalue distribution is broadly similar to the other cases, the low-lying resonance candidates are reproduced less accurately than with the polynomial basis.
Moreover, obtaining results of comparable quality appears to require a more careful tuning of the numerical setup, including the number of basis functions, the Gaussian ranges, and the integration interval used for the interaction matrix.
At the present stage, this makes the trigonometric basis less suitable for a systematic survey.

Taken together, these figures show that the fundamental QNM is reproduced rather stably across different basis choices, while higher-lying states are more sensitive to the basis and to the scaling angle.
This is precisely the behavior expected in a CSM calculation: physical resonance eigenvalues should remain comparatively stable, whereas continuum-related or spurious states tend to move more substantially under changes of the numerical representation.

For convenience, the lowest and next resonance candidates obtained with the different basis sets are summarized in Table~\ref{tab:l2_basis_comparison}.
For the $l=2$ Regge--Wheeler mode, the fundamental frequency is reproduced very stably and is already close to Leaver's value across the basis sets considered here.
By contrast, the second-mode candidate shows a visibly stronger dependence on the basis choice and on the scaling angle.
Among the basis sets examined here, the polynomial basis with $p_{\max}=3$ gives the best overall agreement for both the fundamental mode and the first overtone, while the sin/cos basis reproduces even the fundamental mode less accurately in the present implementation.

It is therefore natural to regard the Polynomial $\times$ real range Gaussian basis as the most practical choice for the subsequent calculations.
The trial run in this subsection thus serves not only as a demonstration of basis dependence, but also as a concrete basis-selection step for the main numerical analysis below.

\begin{table}[ht]
    \centering
    \caption{Comparison of the lowest and next resonance candidates for the $l=2$ Regge--Wheeler mode obtained with different basis sets.
    In this analysis, we set $s=2$ and $M=0.5$.
    For reference, the corresponding Schwarzschild QNM frequencies from Leaver's method are also shown.}
    \label{tab:l2_basis_comparison}
    \begin{tabular}{llcc}
        \toprule
        Basis & Angle & Lowest mode $\omega_1$ & Next mode $\omega_2$ \\
        \midrule
        \multicolumn{4}{l}{\textit{Reference (Leaver)}} \\
        Leaver & --- &
        $0.747343-\rmi0.177925$ &
        $0.693422-\rmi0.547830$ \\
        \midrule
        \multicolumn{4}{l}{\textit{Real range Gaussian basis}} \\
        Real range Gaussian & $\theta=40^\circ$ &
        $0.747343-\rmi0.177958$ &
        $0.750229-\rmi0.481678$ \\
        Real range Gaussian & $\theta=42^\circ$ &
        $0.747349-\rmi0.177940$ &
        $0.738955-\rmi0.497959$ \\
        \midrule
        \multicolumn{4}{l}{\textit{Polynomial $\times$ real range Gaussian basis ($p_{\max}=3$)}} \\
        Polynomial $\times$ Gaussian & $\theta=40^\circ$ &
        $0.747348-\rmi0.177926$ &
        $0.712209-\rmi0.485971$ \\
        Polynomial $\times$ Gaussian & $\theta=42^\circ$ &
        $0.747345-\rmi0.177925$ &
        $0.704895-\rmi0.502870$ \\
        \midrule
        \multicolumn{4}{l}{\textit{sin/cos $\times$ real range Gaussian basis}} \\
        $\sin/\cos \times$ Gaussian & $\theta=40^\circ$ &
        $0.743544-\rmi0.168990$ &
        $0.672161-\rmi0.389585$ \\
        $\sin/\cos \times$ Gaussian & $\theta=42^\circ$ &
        $0.746374-\rmi0.169193 $ &
        $0.673930-\rmi0.403464$ \\
        \bottomrule
    \end{tabular}
\end{table}

\clearpage

\subsection{Basis selection from trial run of more general cases}

Before presenting the main numerical results, we briefly summarize the outcome of a broader trial run aimed at selecting a practical working basis for the subsequent calculations.
The purpose of this subsection is not yet to establish the final numerical accuracy of the method, but rather to identify which class of basis functions captures the low-lying QNM spectrum in a stable and manageable way in the Schwarzschild and extremal Reissner--Nordstr\"om problems.

As candidate trial spaces, we considered several basis families built from real-range Gaussians.
Among them, a particularly natural choice is the Polynomial $\times$ real range Gaussian basis,
\begin{align}
    \phi_{i,n}(r_*)
    \propto
    r_*^n e^{-\alpha_i r_*^2},
    \qquad
    \text{$n=0$, $1$, $2$, \dots} ,
\end{align}
which extends the simplest Gaussian basis by increasing the polynomial degree.
We also tested more explicitly oscillatory trial spaces, such as the sin/cos $\times$ real range Gaussian basis.
The motivation for the latter is clear: since QNM wave functions are oscillatory, one may expect a trigonometric basis to be advantageous.
In practice, however, the situation turned out to be more delicate.

Our exploratory calculations indicate that the polynomial basis is the most practical choice among those examined here.
For the low-lying modes, it reproduces the known QNM frequencies rather well over a moderate range of basis parameters.
By contrast, the sin/cos basis appears to require a more careful tuning of the numerical setup, including the number of basis functions, the Gaussian ranges, and the integration interval used for the interaction matrix.
At the present stage, this makes the trigonometric basis less convenient for a systematic study.

This tendency is already visible in the Schwarzschild benchmark.
The fundamental mode is reproduced stably across different basis parameter sets, whereas the first overtone shows a noticeably larger spread, and still higher modes become increasingly difficult to identify.
This is consistent with the general expectation that relatively narrow resonances are easier to isolate in a finite-basis CSM calculation, while broad and strongly damped modes are more sensitive to basis truncation and to the vicinity of the rotated continuum.

A similar tendency persists in the extremal Reissner--Nordstr\"om case.
The lowest modes can still be identified in reasonable agreement with existing reference values, while higher and broader modes become more ambiguous.
From the present exploratory viewpoint, the important practical conclusion is therefore the following: although several basis choices are in principle possible, the Polynomial $\times$ real range Gaussian basis provides the best balance between numerical stability, flexibility, and ease of implementation at the current stage of the calculation.

For this reason, in the following subsection we adopt the polynomial basis as our main working basis and present the numerical results obtained within that setup.
The trial run described above should thus be understood as a preparatory step, which clarifies both the range of validity and the practical limitations of the present implementation.

\subsection{Numerical results}

Having selected the Polynomial $\times$ real range Gaussian basis as the main working basis, we now present the numerical results obtained with the present CSM formulation.
Our emphasis is on the low-lying QNM frequencies, for which the stability of the calculation is best under control.

\subsubsection{Schwarzschild case}
\begin{figure}[htbp]
 \centering
 \begin{tikzpicture}
  \node at (0,8) 
  {\includegraphics[width=0.45\linewidth]{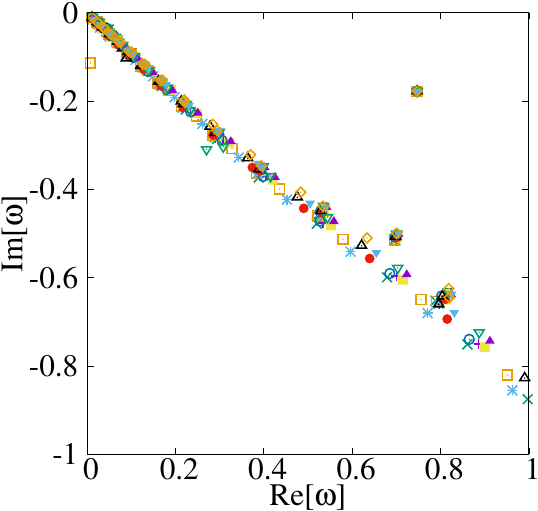}};
  \node at (8,8) 
  {\includegraphics[width=0.45\linewidth]{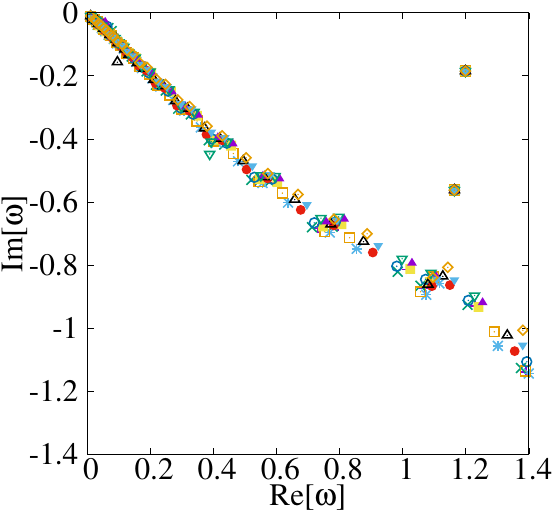}};
  \node at (0,0) 
  {\includegraphics[width=0.45\linewidth]{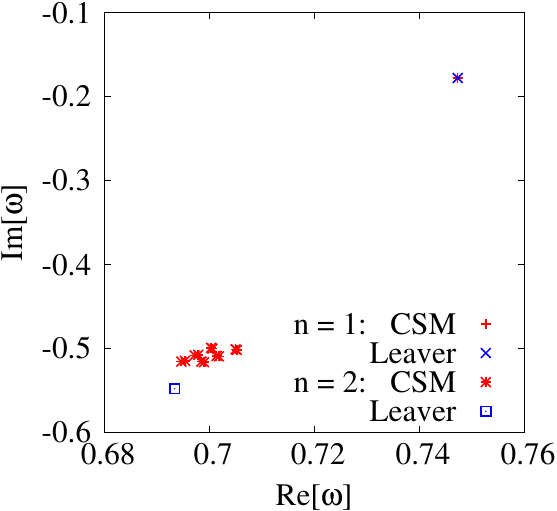}};
  \node at (8,0) 
  {\includegraphics[width=0.45\linewidth]{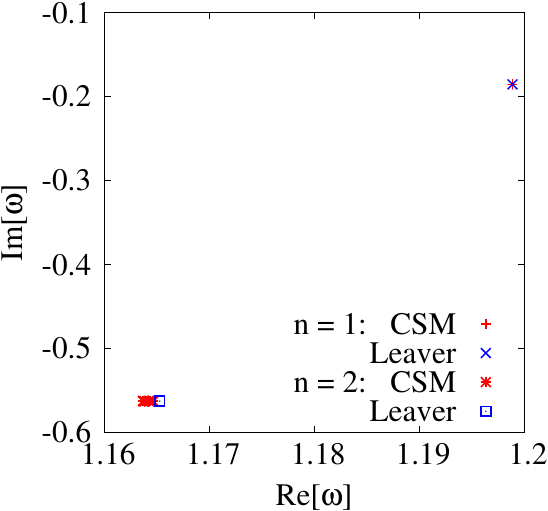}};
  \node[draw, rectangle] at (-1,8) {$l=2$};
  \node[draw, rectangle] at (-1,0) {$l=2$};
  \node[draw, rectangle] at (7,8) {$l=3$};
  \node[draw, rectangle] at (7,0) {$l=3$};
  \draw[thick, red] (2.0625,10.3) circle (0.3);
  \node[above] at (2.0625,10.6) {$n=1$};
  \draw[thick, red] (1.75,8.3) circle (0.3);
  \node[above] at (1.75,8.6) {$n=2$};
  \draw[thick, red] (10.55,10.5) circle (0.3);
  \node[above] at (10.655,10.7) {$n=1$};
  \draw[thick, red] (10.4,8.9) circle (0.3);
  \node[above] at (10.45,9.2) {$n=2$};
 \end{tikzpicture}
 \caption{The Schwarzschild case. $M=0.5$.
 \textbf{Upper panels:} eigenvalues of the complex-scaled Hamiltonian $H_\theta$ for $l=2$ and $l=3$. $\theta=42^\circ$--$44^\circ$.
 \textbf{Lower panels}: comparison of the QNM candidates with Leaver's values.}
 \label{fig:RW_QNM}
\end{figure}
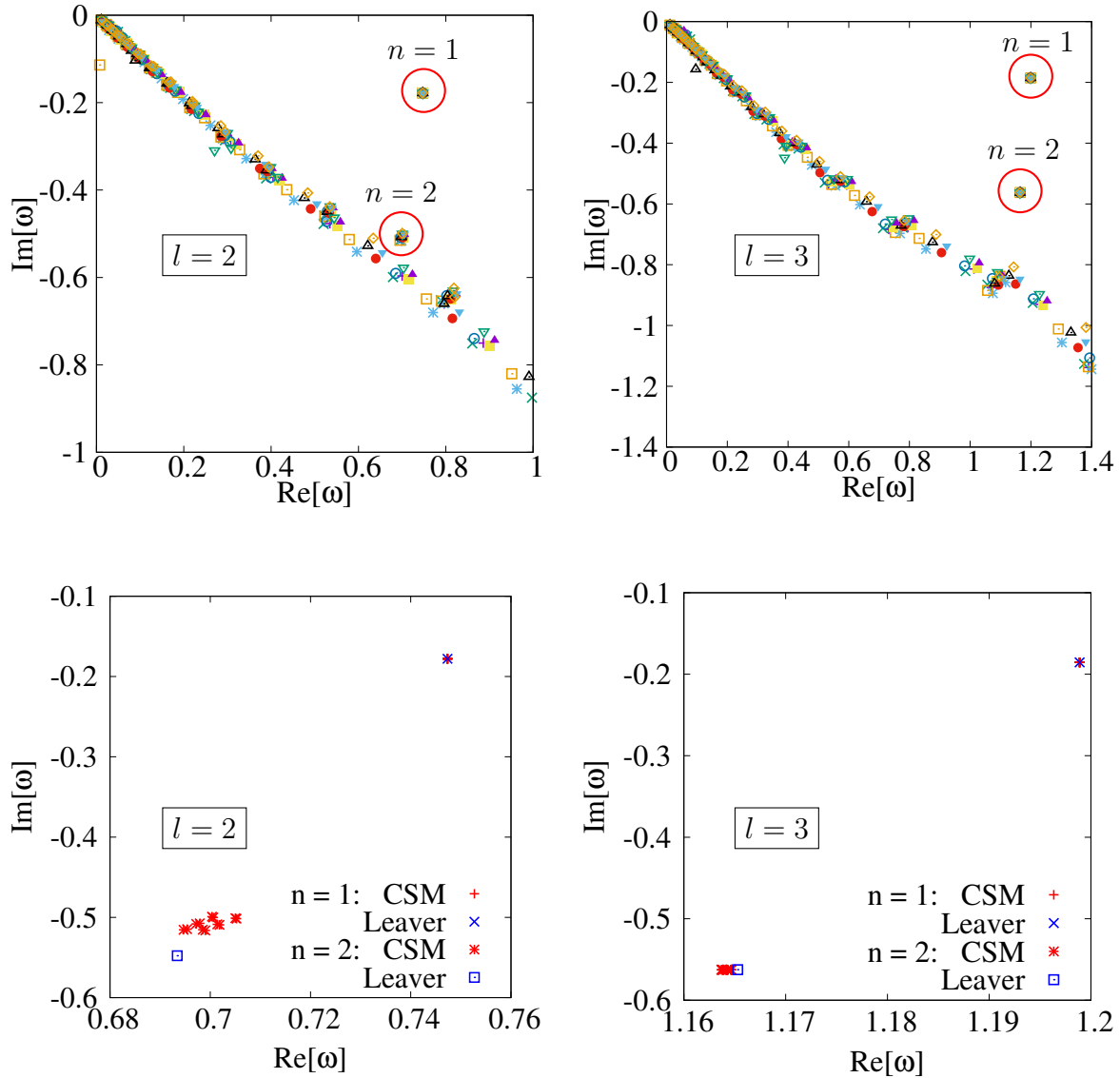

We begin with the Schwarzschild case, which serves as the primary benchmark.
To examine the sensitivity of $\omega$ to Gaussian parameters, we performed calculations for several values of $\theta$ between $42^\circ$ and $44^\circ$ using the four parameter sets listed in Table~\ref{tab:RW_parameters}.
\begin{table}[htbp]
  \centering
  \caption{Parameter sets used to examine the sensitivity of $\omega$ to the Gaussian basis parameters.}
  \label{tab:RW_parameters}
  \begin{tabular}{cccc}
    \toprule
    Set & $i_{\max}$ & $r_{*0}$ & $r_{*\max}$ \\
    \midrule
    1 & 30 & 0.1 & 60 \\
    2 & 30 & 0.1 & 80 \\
    3 & 40 & 0.1 & 60 \\
    4 & 40 & 0.1 & 80 \\
    \bottomrule
  \end{tabular}
\end{table}
The results are shown in Fig.~\ref{fig:RW_QNM}.
The upper panels show the states obtained by diagonalizing $H_\theta$ for $l=2$ and $l=3$. In the lower panels, the values of $\omega$ identified as QNMs are compared with those obtained by Leaver's method for each case.
For the Regge--Wheeler equation, the fundamental mode is reproduced with high accuracy, and the agreement with Leaver's continued-fraction values is very good.
The first overtone can also be identified, although its numerical spread is already noticeably larger than that of the fundamental mode.
For still higher overtones, the identification becomes substantially more difficult within the present implementation, and the extracted values are more sensitive to the basis parameters and the scaling angle.
This behavior is natural in CSM, since highly damped modes lie closer to the rotated continuum and are therefore more vulnerable to finite-basis effects.

\subsubsection{Extremal Reissner--Nordstr\"om case}
We next turn to the extremal Reissner--Nordstr\"om problem.
Here the present framework is particularly useful, since the standard continued-fraction treatment becomes more delicate in the extremal limit.
As in the Schwarzschild case, we examine the sensitivity of the extracted frequencies to the Gaussian basis parameters.
For $(s,l)=(0,0)$, we use the same Gaussian parameter sets as listed in Table~\ref{tab:RW_parameters}. For the other cases, we adopt the parameter sets listed in Table~\ref{tab:exRN_parameters}.
We again performed calculations for several values of $\theta$ from $42^\circ$ to $44^\circ$

\begin{table}[htbp]
  \centering
  \caption{Parameter sets used to examine the sensitivity of $\omega$ to the Gaussian basis parameters.}
  \label{tab:exRN_parameters}
  \begin{tabular}{cccc}
    \toprule
    Set & $i_{\max}$ & $r_{*0}$ & $r_{*\max}$ \\
    \midrule
    I & 40 & 0.1 & 60 \\
    II & 40 & 0.1 & 80 \\
    III & 50 & 0.1 & 60 \\
    IV & 50 & 0.1 & 80 \\
    \bottomrule
  \end{tabular}
\end{table}

\begin{figure}[htbp]
 \centering
 \begin{tikzpicture}
  \node at (0,6) 
  {\includegraphics[width=0.35\linewidth]{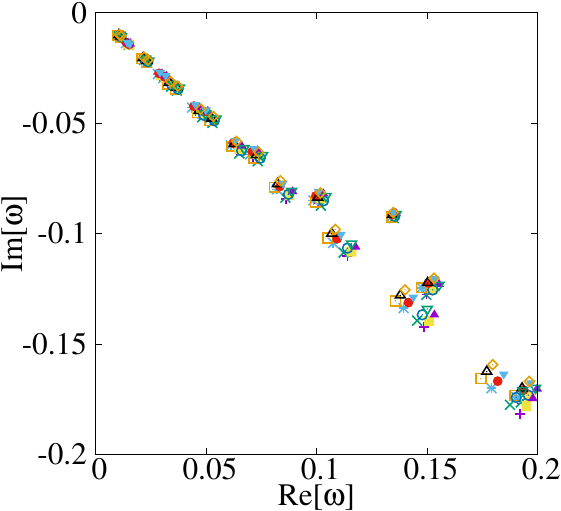}};
  \node at (6,6) 
  {\includegraphics[width=0.35\linewidth]{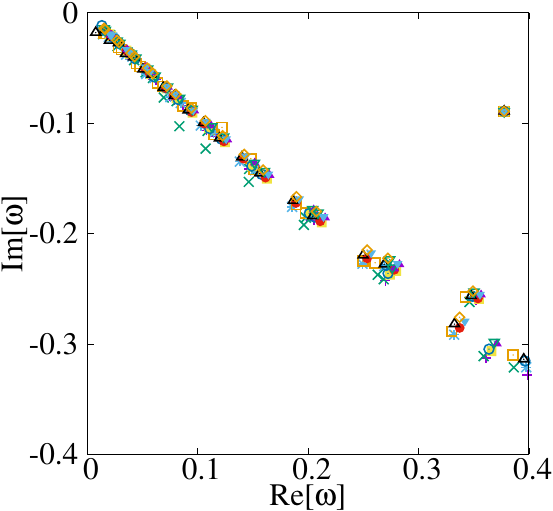}};
  \node at (12,6) 
  {\includegraphics[width=0.35\linewidth]{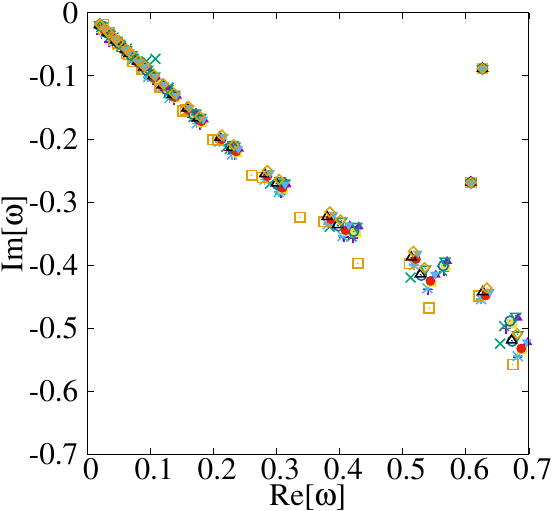}};
  \node at (0,0) 
  {\includegraphics[width=0.35\linewidth]{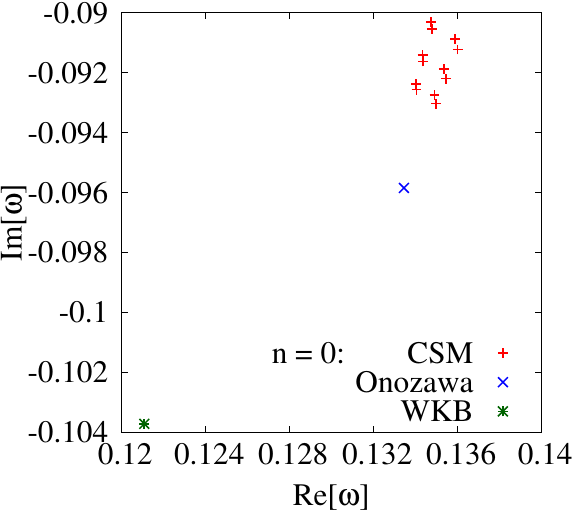}};
  \node at (6,0) 
  {\includegraphics[width=0.35\linewidth]{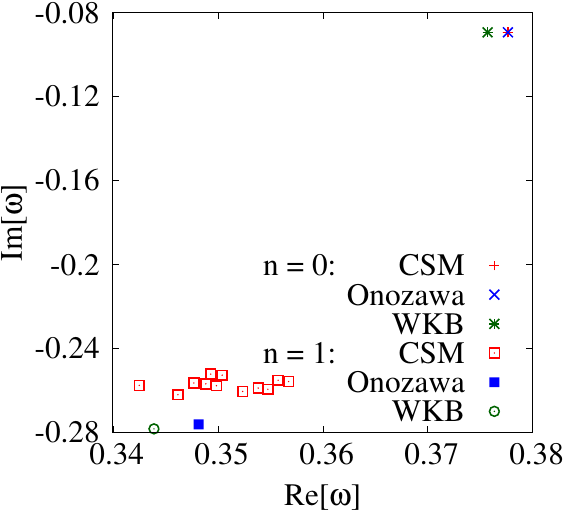}};
  \node at (12,0) 
  {\includegraphics[width=0.35\linewidth]{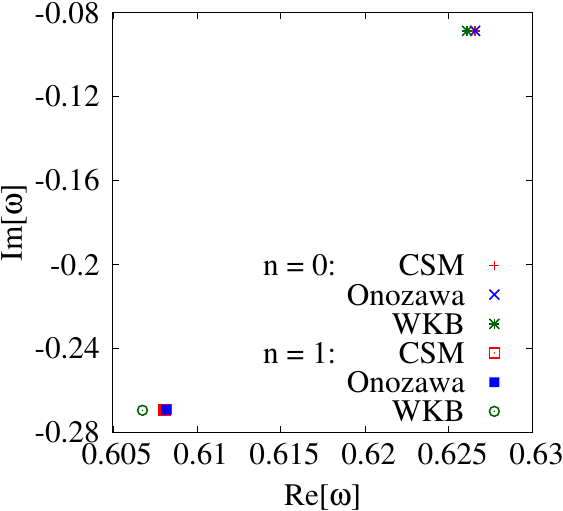}};
  \node[draw, rectangle] at (-1,5) {$l=0$};
  \node[draw, rectangle] at (-0.9,1) {$l=0$};
  \node[draw, rectangle] at (5,5) {$l=1$};
  \node[draw, rectangle] at (5,1) {$l=1$};
  \node[draw, rectangle] at (11,5) {$l=2$};
  \node[draw, rectangle] at (11,1) {$l=2$};
  \draw[thick, red] (1.15,6.4) circle (0.2);
  \node[above] at (1.2,6.7) {$n=0$};
  \draw[thick, red] (8.4,7.5) circle (0.2);
  \node[above] at (8.0,7.8) {$n=0$};
  \draw[thick, red] (8.05,5.55) circle (0.2);
  \node[above] at (8.1,5.9) {$n=1$};  
  \draw[thick, red] (14.15,7.95) circle (0.2);
  \node[above] at (13.3,8.0) {$n=0$};
  \draw[thick, red] (14.05,6.75) circle (0.2);
  \node[above] at (13.5,7.0) {$n=1$};
 \end{tikzpicture}
 \caption{The extremal Reissner--Nordstr\"om scalar case. $M=1$.
 Identification and comparison for $l=0$, $1$ and~$2$ with Onozawa \textit{et al.\/}}
 \label{fig:exRN_scalar_QNM}
\end{figure}
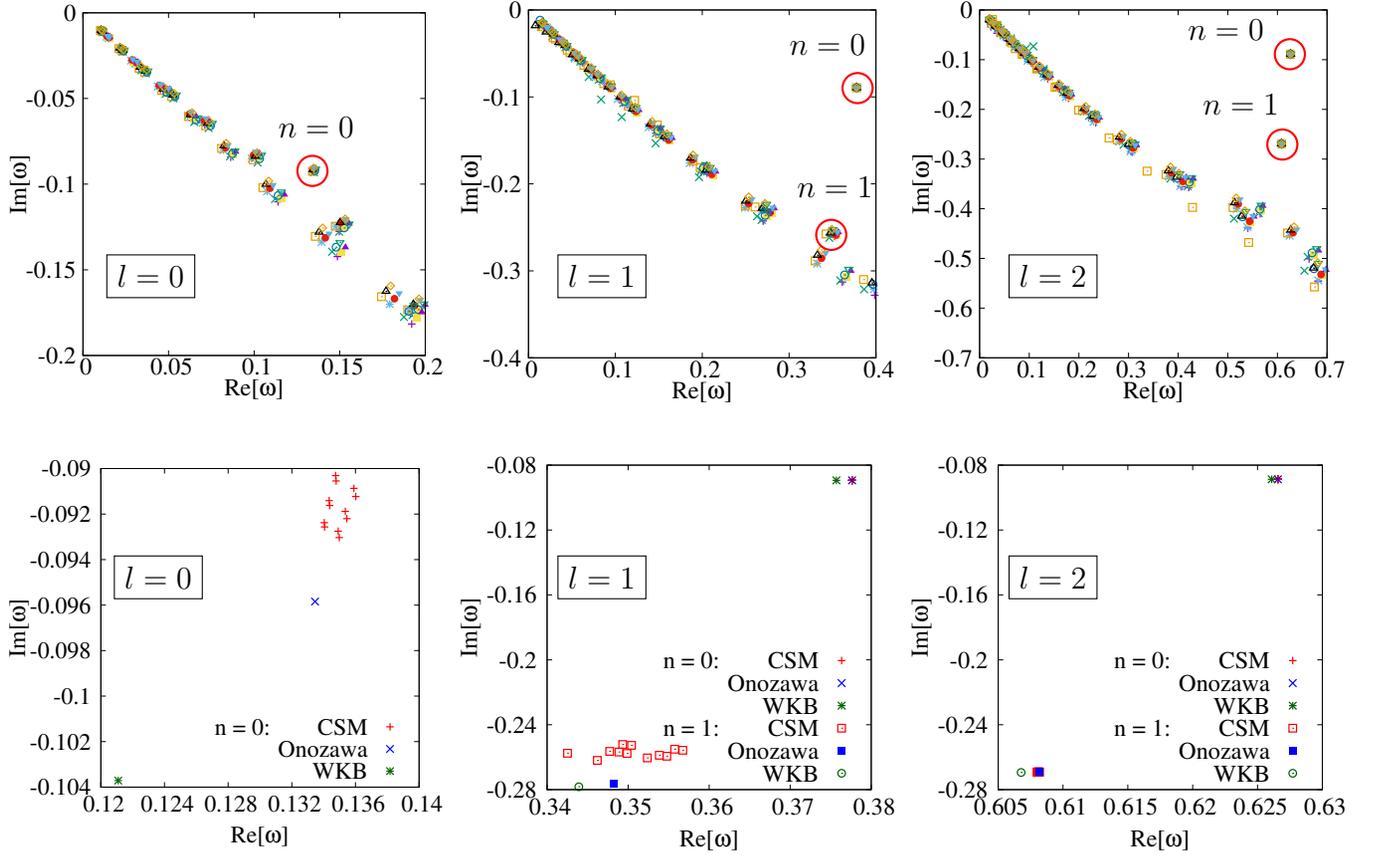

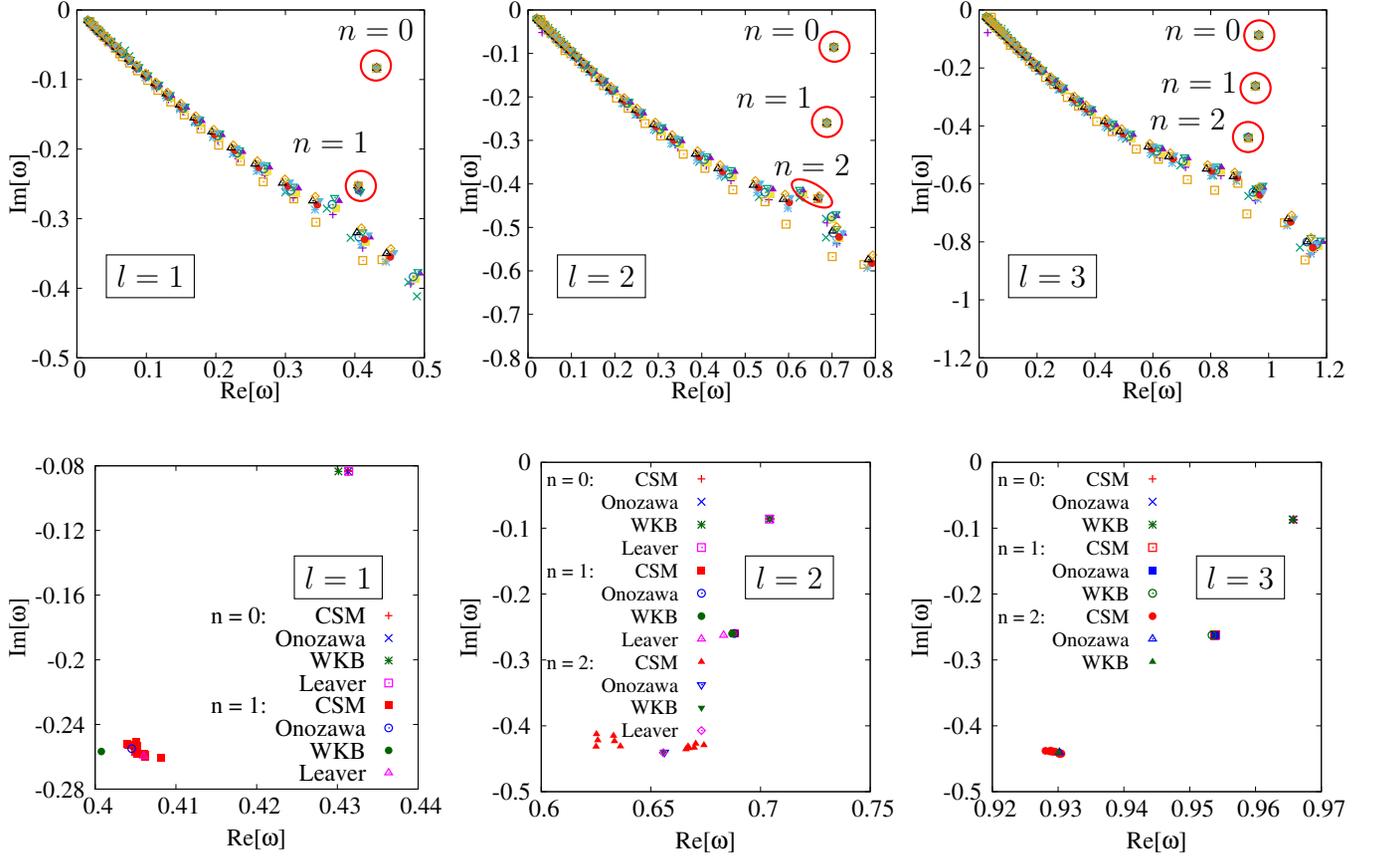
\begin{figure}[htbp]
 \centering
 \begin{tikzpicture}
  \node at (0,6) 
  {\includegraphics[width=0.35\linewidth]{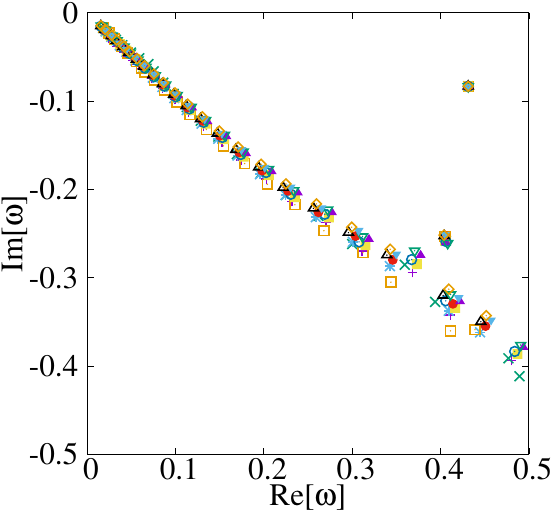}};
  \node at (6,6) 
  {\includegraphics[width=0.35\linewidth]{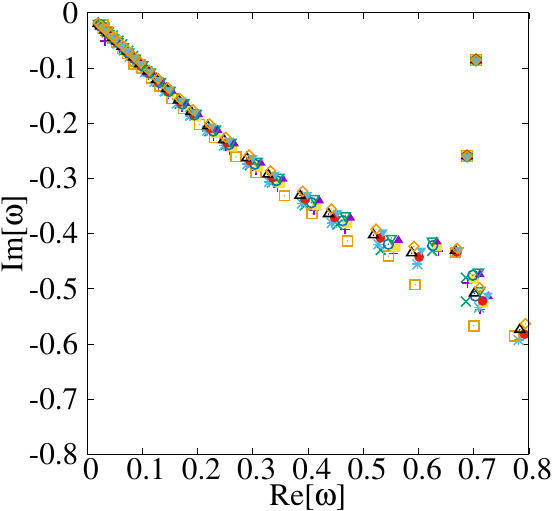}};
  \node at (12,6) 
  {\includegraphics[width=0.35\linewidth]{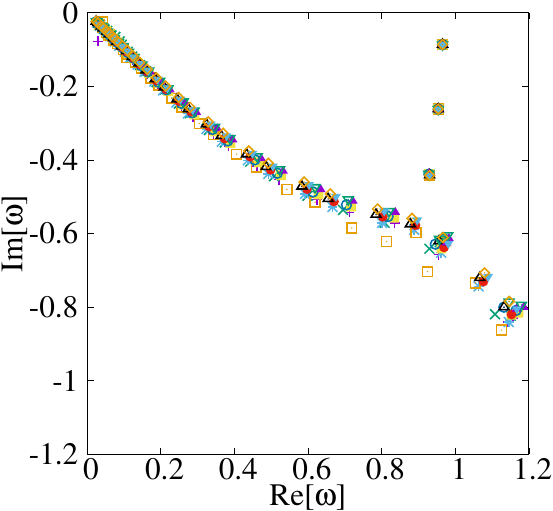}};
  \node at (0,0) 
  {\includegraphics[width=0.35\linewidth]{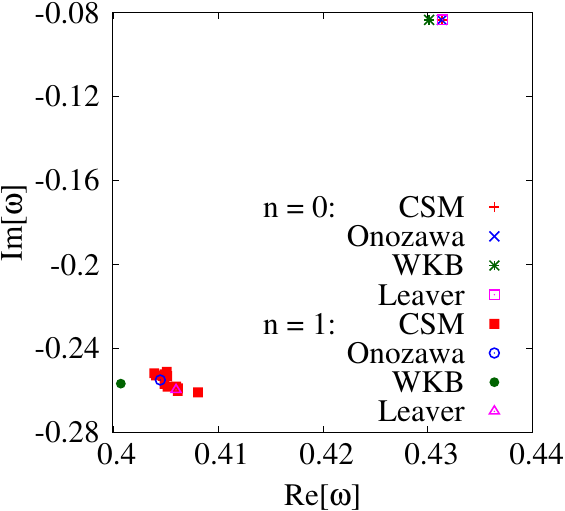}};
  \node at (6,0) 
  {\includegraphics[width=0.35\linewidth]{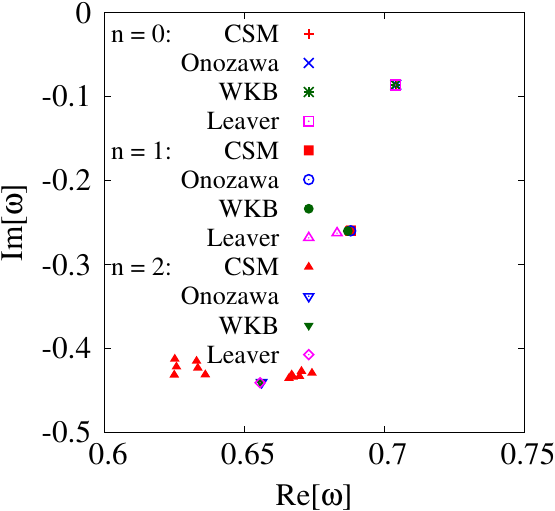}};
  \node at (12,0) 
  {\includegraphics[width=0.35\linewidth]{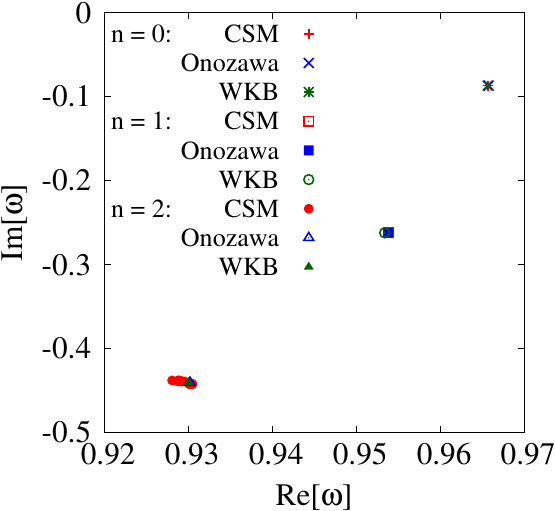}};
  \node[draw, rectangle] at (-1,5) {$l=1$};
  \node[draw, rectangle] at (1.5,1) {$l=1$};
  \node[draw, rectangle] at (5,5) {$l=2$};
  \node[draw, rectangle] at (7.5,1) {$l=2$};
  \node[draw, rectangle] at (11,5) {$l=3$};
  \node[draw, rectangle] at (13.5,1) {$l=3$};
  \draw[thick, red] (2.0,7.8) circle (0.2);
  \node[above] at (2.0,8.0) {$n=0$};
  \draw[thick, red] (1.8,6.2) circle (0.2);
  \node[above] at (1.4,6.5) {$n=1$};
  \draw[thick, red] (8.1,8.05) circle (0.2);
  \node[above] at (7.4,8.0) {$n=0$};
  \draw[thick, red] (8.0,7.05) circle (0.2);
  \node[above] at (7.3,7.1) {$n=1$};  
  \draw[thick, red] (7.8,6.1) ellipse [x radius=0.3, y radius=0.13, rotate=-30];
  \node[above] at (7.8,6.2) {$n=2$};  
  \draw[thick, red] (13.75,8.20) circle (0.2);
  \node[above] at (13.0,8.0) {$n=0$};
  \draw[thick, red] (13.7,7.5) circle (0.2);
  \node[above] at (12.95,7.3) {$n=1$};
  \draw[thick, red] (13.6,6.85) circle (0.2);
  \node[above] at (12.8,6.8) {$n=2$};
 \end{tikzpicture}
 \caption{The extremal Reissner--Nordstr\"om odd-parity electromagnetic case. $M=1$.}
 \label{fig:exRN_em_QNM}
\end{figure}

\begin{figure}[htbp]
 \centering
 \begin{tikzpicture}
  \node at (0,6) 
  {\includegraphics[width=0.35\linewidth]{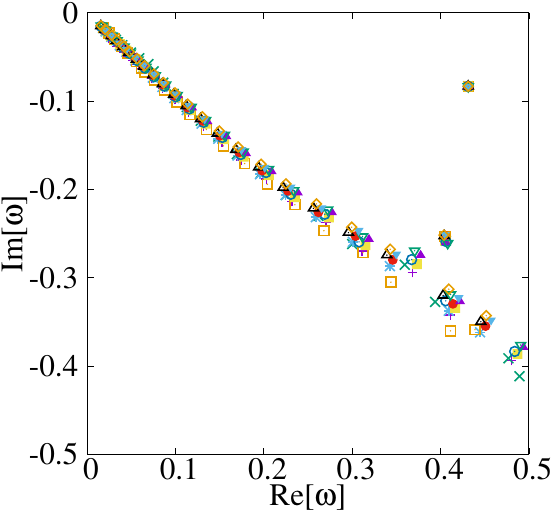}};
  \node at (6,6) 
  {\includegraphics[width=0.35\linewidth]{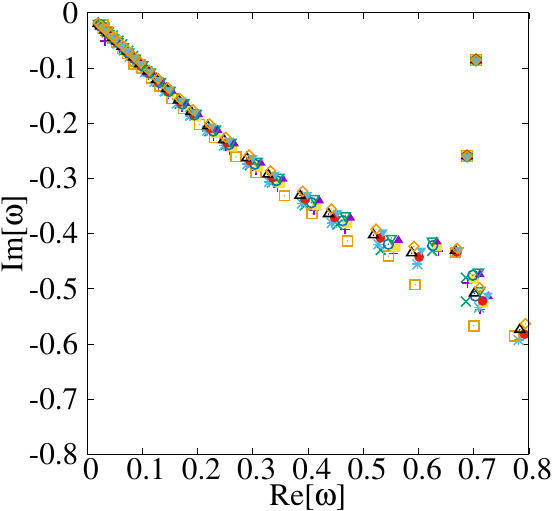}};
  \node at (12,6) 
  {\includegraphics[width=0.35\linewidth]{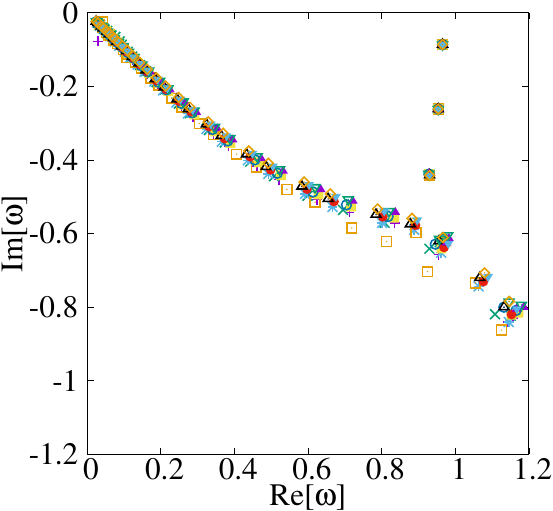}};
  \node at (0,0) 
  {\includegraphics[width=0.35\linewidth]{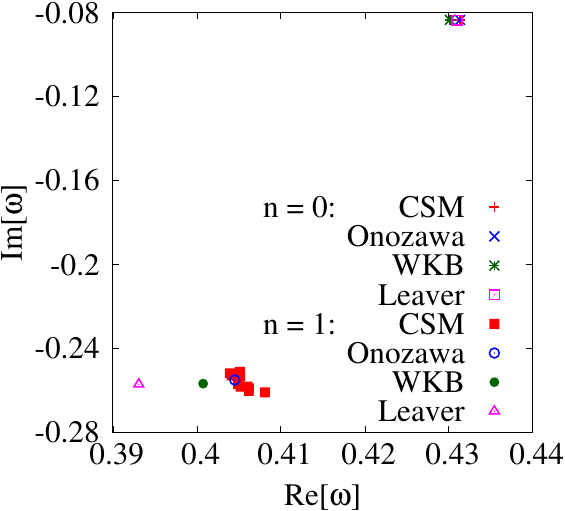}};
  \node at (6,0) 
  {\includegraphics[width=0.35\linewidth]{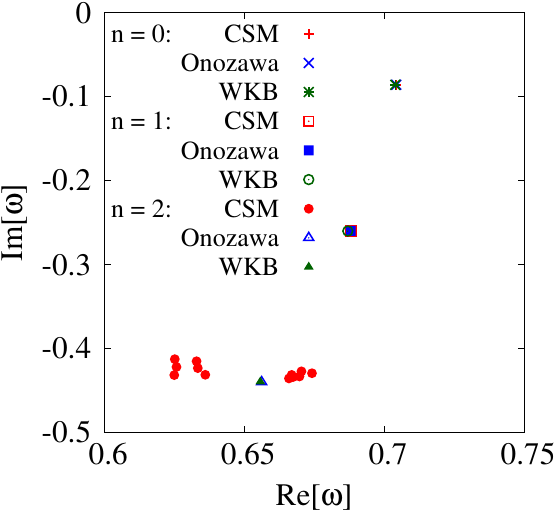}};
  \node at (12,0) 
  {\includegraphics[width=0.35\linewidth]{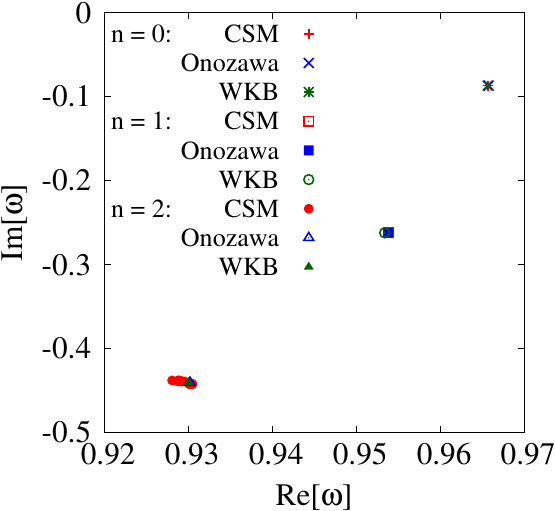}};
  \node[draw, rectangle] at (-1,5) {$l=2$};
  \node[draw, rectangle] at (1.5,1) {$l=2$};
  \node[draw, rectangle] at (5,5) {$l=3$};
  \node[draw, rectangle] at (7.5,1) {$l=3$};
  \node[draw, rectangle] at (11,5) {$l=4$};
  \node[draw, rectangle] at (13.5,1) {$l=4$};
  \draw[thick, red] (2.0,7.8) circle (0.2);
  \node[above] at (2.0,8.0) {$n=0$};
  \draw[thick, red] (1.8,6.2) circle (0.2);
  \node[above] at (1.4,6.5) {$n=1$};
  \draw[thick, red] (8.1,8.05) circle (0.2);
  \node[above] at (7.4,8.0) {$n=0$};
  \draw[thick, red] (8.0,7.05) circle (0.2);
  \node[above] at (7.3,7.1) {$n=1$};  
  \draw[thick, red] (7.8,6.1) ellipse [x radius=0.3, y radius=0.13, rotate=-30];
  \node[above] at (7.8,6.2) {$n=2$};  
  \draw[thick, red] (13.75,8.20) circle (0.2);
  \node[above] at (13.0,8.0) {$n=0$};
  \draw[thick, red] (13.7,7.5) circle (0.2);
  \node[above] at (12.95,7.3) {$n=1$};
  \draw[thick, red] (13.6,6.85) circle (0.2);
  \node[above] at (12.8,6.8) {$n=2$};
 \end{tikzpicture}
 \caption{The extremal Reissner--Nordstr\"om odd-parity gravitational case. $M=1$.}
 \label{fig:exRN_grav_QNM}
\end{figure}

The numerical results are shown in Figs.~\ref{fig:exRN_scalar_QNM}, \ref{fig:exRN_em_QNM}, and~\ref{fig:exRN_grav_QNM} for scalar ($s=0$), odd-parity electromagnetic ($s=1$), and odd-parity gravitational ($s=2$) perturbations, respectively.
As in the Schwarzschild case, the upper panels show the eigenvalues obtained by diagonalizing the complex-scaled Hamiltonian $H_\theta$ for several values of $l$, while the lower panels compare the frequencies identified as QNMs with existing reference values.
In the extremal Reissner--Nordstr\"om case, the lower panels include the results of Onozawa \textit{et al.\/}~\cite{Onozawa:1995vu} for comparison.

For the perturbations considered here, the low-lying QNM frequencies are again reproduced in reasonable agreement with the existing calculations.
The agreement is especially good for the lowest and relatively narrow modes, whereas broader and more strongly damped modes exhibit a larger spread under changes of the basis parameters and the scaling angle.
This is qualitatively the same tendency as in the Schwarzschild benchmark and is consistent with the general expectation that highly damped modes, being closer to the rotated continuum, are more sensitive to finite-basis effects in the CSM.

It is also worth noting that, whenever both Onozawa \textit{et al.\/} and WKB values are available, the present CSM results tend in general to lie closer to the former than to the latter.
From the viewpoint of resonance identification, this is an encouraging sign that the present implementation is correctly capturing the low-lying QNM spectrum in the extremal Reissner--Nordstr\"om background.

Another useful consistency check is provided by the known isospectrality between the odd-parity electromagnetic and odd-parity gravitational sectors in extremal Reissner--Nordstr\"om spacetime.
Within the numerical precision of the present calculation, we confirm that the same QNM frequencies are obtained in these two sectors.
This agreement provides an additional indication that the present CSM implementation is correctly capturing the low-lying resonance spectrum.

\subsubsection{Summary of low-lying QNM frequencies}
Overall, the numerical results support the following picture.
For relatively narrow, isolated modes, the present CSM formulation performs well and yields QNM frequencies in good agreement with established methods.
For broader, more strongly damped modes, however, numerical stability is reduced, and a more careful analysis of convergence with respect to basis size, basis parameters, integration range, and scaling angle will be required.
Accordingly, the present results should be regarded as a first systematic demonstration of the method rather than as its final optimized form.

In the remainder of this section, we summarize the low-lying frequencies obtained for the Schwarzschild and extremal Reissner--Nordstr\"om cases, and compare them with the available reference values reviewed in Appendix~\ref{sec:leaver}.

Table~\ref{tab:result_schw} presents the numerical results obtained by the CSM for the Schwarzschild case.
\revone{To characterize the sensitivity of the extracted
QNM frequencies to the sampled numerical parameter sets, we use the
componentwise parameter-variation spreads returned by the automated
procedure of Sec.~\ref{sec:stability_qnm}.}
Let $\omega_i$ denote the frequency obtained in the $i$th run, and let $\vsub{N}{samp}$ be the total number of runs.
We then define the average frequency by
\begin{align}
    \bar{\omega}
    =
    \frac{1}{\vsub{N}{samp}}
    \sum_{i=1}^{\vsub{N}{samp}}
    \omega_i ,
\end{align}
and \revone{characterize the sensitivity to the tested numerical parameter sets by
the componentwise parameter-variation spreads
\begin{align}
    \Delta \omega_R
    =
    \max_i
    \left|
        \re\omega_i-\re\bar{\omega}
    \right| ,
    \qquad
    \Delta \omega_I
    =
    \max_i
    \left|
        \im\omega_i-\im\bar{\omega}
    \right|.
\end{align}
Accordingly, we quote the final result as
\begin{align}
    \re\omega = \re\bar{\omega} \pm \Delta\omega_R , \qquad
    \im\omega = \im\bar{\omega} \pm \Delta\omega_I.
\end{align}
Here, $\Delta\omega_{R,I}$} should be understood as a practical measure of numerical stability rather than as a statistical error bar.
\revone{In the tables, an entry
\(x(a)-\rmi y(b)\) denotes the averaged frequency
\(\re\bar\omega=x\), \(-\im\bar\omega=y\), together with the
corresponding componentwise spreads \(a\) and \(b\). These quantities are
stability diagnostics over the sampled parameter sets and should not be
interpreted as statistically defined error bars or as rigorous estimates of
the finite-basis truncation error.}

\begin{table}[htbp]
\centering
\caption{Numerical results of QNM frequencies of Schwarzschild.
\revone{The numbers in parentheses indicate the maximum componentwise spread under the specified variations of the scaling angle and basis parameters. They are stability diagnostics and should not be interpreted as statistically defined or rigorously converged error bars.
An unmarked CSM entry denotes a robustly identified mode, while a dagger denotes a tentative candidate according to the reference-free procedure of Sec.~\ref{sec:stability_qnm}. The phrase ``not identified'' denotes a marginal feature suggestive of
a candidate that did not satisfy the isolation and cross-angle
persistence criteria strongly enough for a frequency to be quoted.
A dash indicates that no CSM frequency is quoted for that entry in the
present scan. Neither ``not identified'' nor a dash should be
interpreted as evidence that the corresponding physical QNM is absent.}}
\label{tab:result_schw}
\begin{tabular}{cl
                l@{$-\rmi$}lc
                l@{$-\rmi$}l}
\toprule
$n$ & Method & \multicolumn{2}{c}{$l=2$} &  & \multicolumn{2}{c}{$l=3$} \\
\midrule 
$1$ & CSM    & $0.7473434(10)$ & $0.17792461(83)$
    &        & $1.19888660(11)$ & $0.18540611(13)$ \\
    & Leaver & $0.747343$    & $0.177925$
    &        & $1.198887$    & $0.185406$ \\
\addlinespace
$2$ & CSM    & $0.6998(55)$ & $0.5081(90)^\dagger$
    &        & $1.16423(53)$  & $0.56265(27)$ \\
    & Leaver & $0.693422$   & $0.547830$
    &        & $1.165288$   & $0.562596$ \\
\addlinespace
$3$ & CSM    & \multicolumn{2}{c}{---}
    &        & \multicolumn{2}{c}{not identified} \\
    & Leaver & $0.602107$ & $0.956554$
    &        & $1.103370$ & $0.958186$ \\
\bottomrule
\end{tabular}
\end{table}

Tables~\ref{tab:result_rn0}--\ref{tab:result_rn0_2}, \ref{tab:result_rn1}--\ref{tab:result_rn1_2}, and~\ref{tab:result_rn2}--\ref{tab:result_rn2_2} show the corresponding results for the extremal Reissner--Nordstr\"om case with scalar ($s=0$), odd-parity electromagnetic ($s=1$), and odd-parity gravitational ($s=2$) perturbations, respectively.
Particular attention will be paid to the degree of stability under changes of the basis parameters, as well as to the distinction between robust resonance eigenvalues and basis-dependent artifacts.

\begin{table}[htbp]
\centering
\caption{QNMs of extremal Reissner--Nordstr\"om scalar ($s=0$) for $l=0$ and $l=1$}
\label{tab:result_rn0}
\begin{tabular}{l
                l@{$-\rmi$}l
                l@{$-\rmi$}l}
\toprule
Method & \multicolumn{2}{c}{$l=0$} & \multicolumn{2}{c}{$l=1$} \\
\midrule
\multicolumn{5}{c}{$n=0$} \\
CSM     & $0.1349(11)$  & $0.0917(14)$
        & $0.3776424(14)$ & $0.08938386(76)$ \\
Onozawa & $0.13346$     & $0.095844$
        & $0.37764$     & $0.089384$ \\
WKB     & $0.12109$     & $0.10371$
        & $0.37570$     & $0.08936$ \\
\addlinespace
\multicolumn{5}{c}{$n=1$} \\
CSM     & \multicolumn{2}{c}{---}
        & $0.3507(82)$ & $0.2571(49)^\dagger$ \\
Onozawa & $0.092965$    & $0.33065$
        & $0.34818$     & $0.27614$ \\
WKB     & $0.09157$     & $0.33742$
        & $0.34392$     & $0.27828$ \\
\addlinespace
\multicolumn{5}{c}{$n=2$} \\
CSM     & \multicolumn{2}{c}{---}
        & \multicolumn{2}{c}{---} \\
Onozawa & $0.075081$ & $0.58833$
        & $0.29846$  & $0.48643$ \\
WKB     & $0.05056$  & $0.57164$
        & $0.29661$  & $0.48145$ \\
\bottomrule
\end{tabular}
\end{table}

\begin{table}[htbp]
\centering
\caption{QNMs of extremal Reissner--Nordstr\"om scalar ($s=0$) for $l=2$}
\label{tab:result_rn0_2}
\begin{tabular}{l
                l@{$-\rmi$}l}
\toprule
Method & \multicolumn{2}{c}{$l=2$} \\
\midrule
\multicolumn{3}{c}{$n=0$} \\
CSM     & $0.62657272(19)$ & $0.088748359(98)$ \\
Onozawa & $0.62657$     & $0.088748$ \\
WKB     & $0.62609$     & $0.08873$ \\
\addlinespace
\multicolumn{3}{c}{$n=1$} \\
CSM     & $0.60810(10)$   & $0.26920(11)$ \\
Onozawa & $0.60817$     & $0.26909$ \\
WKB     & $0.60677$     & $0.26944$ \\
\addlinespace
\multicolumn{3}{c}{$n=2$} \\
CSM     & \multicolumn{2}{c}{not identified} \\
Onozawa & $0.57287$  & $0.45820$ \\
WKB     & $0.57254$  & $0.45750$ \\
\bottomrule
\end{tabular}
\end{table}

\begin{table}[htbp]
\centering
\caption{QNMs of extremal Reissner--Nordstr\"om odd-parity electromagnetic ($s=1$) for $l=1$ and $l=2$}
\label{tab:result_rn1}
\begin{tabular}{l
                l@{$-\rmi$}l
                l@{$-\rmi$}l}
\toprule
Method & \multicolumn{2}{c}{$l=1$} & \multicolumn{2}{c}{$l=2$} \\
\midrule
\multicolumn{5}{c}{$n=0$} \\
CSM     & $0.431340738(82)$ & $0.08346031(23)$
        & $0.704304022(14)$ & $0.085973391(76)$ \\
Onozawa & $0.43134$     & $0.083460$
        & $0.70430$     & $0.085973$ \\
Leaver  & $0.431415$    & $0.08343$
        & $0.704075$    & $0.086205$ \\
WKB     & $0.43013$     & $0.08349$
        & $0.70398$     & $0.08596$ \\
\addlinespace
\multicolumn{5}{c}{$n=1$} \\
CSM     & $0.4054(27)$ & $0.2556(53)$
        & $0.688059(56)$  & $0.259911(49)$ \\
Onozawa & $0.40452$    & $0.25498$
        & $0.68804$    & $0.25992$ \\
Leaver  & $0.40602$    & $0.259705$
        & $0.68315$    & $0.26256$ \\
WKB     & $0.40076$    & $0.25675$
        & $0.68701$    & $0.26014$ \\
\addlinespace
\multicolumn{5}{c}{$n=2$} \\
CSM     & \multicolumn{2}{c}{---}
        & $0.649(25)$ & $0.427(15)^\dagger$ \\
Onozawa & $0.35340$    & $0.44137$
        & $0.65624$    & $0.44007$ \\
Leaver  & $0.35347$    & $0.44260$
        & $0.655675$   & $0.4408$ \\
WKB     & $0.35136$    & $0.44210$
        & $0.65575$    & $0.43986$ \\
\bottomrule
\end{tabular}
\end{table}

\begin{table}[htbp]
\centering
\caption{QNMs of extremal Reissner--Nordstr\"om odd-parity electromagnetic ($s=1$) for $l=3$}
\label{tab:result_rn1_2}
\begin{tabular}{l
                l@{$-\rmi$}l}
\toprule
Method & \multicolumn{2}{c}{$l=3$} \\
\midrule
\multicolumn{3}{c}{$n=0$} \\
CSM     & $0.965762603(31)$ & $0.087001340(45)$ \\
Onozawa & $0.96576$     & $0.087001$ \\
Leaver  & \multicolumn{2}{c}{---} \\
WKB     & $0.96563$     & $0.08700$ \\
\addlinespace
\multicolumn{3}{c}{$n=1$} \\
CSM     & $0.953816(10)$ & $0.2621206(52)$ \\
Onozawa & $0.95381$   & $0.26212$ \\
Leaver  & \multicolumn{2}{c}{---} \\
WKB     & $0.95339$   & $0.26218$ \\
\addlinespace
\multicolumn{3}{c}{$n=2$} \\
CSM     & $0.9296(16)$ & $0.4403(26)$ \\
Onozawa & $0.93020$    & $0.44064$ \\
Leaver  & \multicolumn{2}{c}{---} \\
WKB     & $0.93004$    & $0.44056$ \\
\bottomrule
\end{tabular}
\end{table}

\begin{table}[htbp]
\centering
\caption{QNMs of extremal Reissner--Nordstr\"om odd-parity gravitational ($s=2$) for $l=2$ and $l=3$}
\label{tab:result_rn2}
\begin{tabular}{l
                l@{$-\rmi$}l
                l@{$-\rmi$}l}
\toprule
Method & \multicolumn{2}{c}{$l=2$} & \multicolumn{2}{c}{$l=3$} \\
\midrule
\multicolumn{5}{c}{$n=0$} \\
CSM     & $0.431340738(82)$ & $0.08346031(23)$
        & $0.704304022(14)$ & $0.085973391(76)$ \\
Onozawa & $0.43134$     & $0.083460$
        & $0.70430$     & $0.085973$ \\
Leaver  & $0.431415$    & $0.08343$
        & $0.704075$    & $0.086205$ \\
WKB     & $0.43013$     & $0.08349$
        & $0.70398$     & $0.08596$ \\
\addlinespace
\multicolumn{5}{c}{$n=1$} \\
CSM     & $0.4054(27)$ & $0.2556(53)$
        & $0.688059(56)$  & $0.259911(49)$ \\
Onozawa & $0.40452$    & $0.25498$
        & $0.68804$    & $0.25992$ \\
Leaver  & $0.40602$    & $0.259705$
        & $0.68315$    & $0.26256$ \\
WKB     & $0.40076$    & $0.25675$
        & $0.68701$    & $0.26014$ \\
\addlinespace
\multicolumn{5}{c}{$n=2$} \\
CSM     & \multicolumn{2}{c}{---}
        & $0.649(25)$ & $0.427(15)^\dagger$ \\
Onozawa & $0.35340$    & $0.44137$
        & $0.65624$    & $0.44007$ \\
Leaver  & $0.35347$    & $0.44260$
        & $0.655675$   & $0.4408$ \\
WKB     & $0.35136$    & $0.44210$
        & $0.65575$    & $0.43986$ \\
\bottomrule
\end{tabular}
\end{table}

\begin{table}[htbp]
\centering
\caption{QNMs of extremal Reissner--Nordstr\"om odd-parity gravitational ($s=2$) for $l=4$}
\label{tab:result_rn2_2}
\begin{tabular}{l
                l@{$-\rmi$}l}
\toprule
Method & \multicolumn{2}{c}{$l=4$} \\
\midrule
\multicolumn{3}{c}{$n=0$} \\
CSM     & $0.965762603(31)$ & $0.087001340(45)$ \\
Onozawa & $0.96576$     & $0.087001$ \\
Leaver  & \multicolumn{2}{c}{---} \\
WKB     & $0.96563$     & $0.08700$ \\
\addlinespace
\multicolumn{3}{c}{$n=1$} \\
CSM     & $0.953816(10)$ & $0.2621206(52)$ \\
Onozawa & $0.95381$   & $0.26212$ \\
Leaver  & \multicolumn{2}{c}{---} \\
WKB     & $0.95339$   & $0.26218$ \\
\addlinespace
\multicolumn{3}{c}{$n=2$} \\
CSM     & $0.9296(16)$ & $0.4403(26)$ \\
Onozawa & $0.93020$    & $0.44064$ \\
Leaver  & \multicolumn{2}{c}{---} \\
WKB     & $0.93004$    & $0.44056$ \\
\bottomrule
\end{tabular}
\end{table}

A few general trends may be read off from Tables~\ref{tab:result_schw}--\ref{tab:result_rn2_2}.
In both the Schwarzschild and extremal Reissner--Nordstr\"om cases, the agreement with the reference values is best for the lowest and relatively narrow modes.
The fundamental mode is reproduced very accurately, whereas the numerical spread becomes larger for higher overtones.
This is consistent with the general expectation in the CSM that strongly damped modes lie closer to the rotated continuum and are therefore more sensitive to finite-basis effects.

In the extremal Reissner--Nordstr\"om case, the present CSM values tend in general to be closer to the results of Onozawa \textit{et al.\/} than to the corresponding WKB values.
Moreover, the odd-parity electromagnetic and odd-parity gravitational sectors yield the same low-lying frequencies within the present numerical precision, in agreement with the known isospectrality of these sectors in the extremal Reissner--Nordstr\"om background.
Taken together, these results indicate that the present implementation already captures the robust low-lying resonance spectrum reliably, while broader and more highly damped modes still require a more systematic study of convergence and numerical stability.

\revone{The numerical accuracy is strongly mode dependent. For narrow and
well-isolated modes, the present calculation reproduces five or more
significant digits of the available benchmark frequencies. For broader
modes close to the rotated continuum, however, the deviations can remain
at the percent level and the parameter-variation spread does not constitute
a rigorous estimate of the truncation error. We therefore do not assign a
uniform accuracy or convergence rate to the method. In the standard
Schwarzschild problem, the optimized continued-fraction method remains
both more accurate and more efficient for computing individual
frequencies.}

\section{Conclusion}
\label{sec:conclusion}

In this paper, we studied black-hole quasinormal modes within the complex scaling method (CSM).
Our main purpose was to develop a common CSM-based framework for the Schwarzschild and Reissner--Nordstr\"om families, and to compute their QNM frequencies in a way that remains close to the resonance-theoretic structure of the problem.

We first reviewed the Regge--Wheeler and Zerilli--Moncrief formulations and then recast the QNM problem as a non-Hermitian spectral problem by means of complex scaling.
This provides a framework in which outgoing resonances can be treated as discrete spectral data of the complex-scaled operator, rather than being imposed only indirectly through the original boundary conditions.
In this sense, the present approach is intended not merely as another numerical scheme, but as a formulation of black-hole QNMs as resonances in a controlled spectral setting.

Our immediate benchmark in this work was the Schwarzschild problem, while the broader target was the Reissner--Nordstr\"om family, including the approach to extremality.
A particular motivation for this extension is that standard Leaver-type methods become more delicate in the extremal limit, where the naive Frobenius structure at the horizon is lost.
The CSM framework is attractive precisely because it is less directly tied to such special analytic structures.

\revone{For the standard Schwarzschild benchmark, the present CSM implementation is not intended to replace Leaver's continued-fraction method in terms of raw numerical precision or computational efficiency. Its principal advantage is structural: the outgoing resonance problem is represented as a non-Hermitian $L^2$ eigenvalue problem without relying on a problem-specific Frobenius recurrence, and the resonance poles and the rotated continuum are encoded in the same spectral operator. This feature is potentially useful in problems with altered singularity structures, coupled-channel potentials, or parameter-dependent resonance trajectories. The method should therefore be regarded as complementary to continued-fraction, time-domain, and hyperboloidal spectral approaches rather than as universally superior to them.}

\revone{Another practical advantage is that CSM does not require the
outgoing asymptotic solution to be imposed explicitly. It can
therefore accommodate Coulomb-like long-range potentials, for which
the asymptotic solutions are Coulomb functions rather than plane
waves, provided that the potential admits the analytic continuation
required by complex scaling.}

At the numerical level, our trial calculations indicate that the Polynomial $\times$ real range Gaussian basis provides the most practical balance between stability and flexibility among the basis sets examined here.
Using this basis, we found that the low-lying QNM frequencies are reproduced well in both the Schwarzschild and extremal Reissner--Nordstr\"om cases.
In particular, the Schwarzschild fundamental mode is obtained in very good agreement with Leaver's values, while higher overtones show a larger spread
under changes of the basis parameters and the scaling angle.
This tendency is consistent with the general expectation in the CSM that broad and strongly damped modes, being closer to the rotated continuum, are more sensitive to finite-basis effects.

For the extremal Reissner--Nordstr\"om case, the low-lying scalar, odd-parity electromagnetic, and odd-parity gravitational modes are again reproduced in reasonable agreement with existing calculations.
Whenever both Onozawa \textit{et al.\/} and WKB values are available, the present CSM results tend in general to lie closer to the former than to the latter.
We also confirmed, within the present numerical precision, the known isospectrality between the odd-parity electromagnetic and odd-parity gravitational sectors in the extremal Reissner--Nordstr\"om background.
Taken together, these results indicate that the present implementation already captures the robust low-lying resonance spectrum reliably.

At the same time, the present numerical study should still be regarded as preliminary in some respects.
For broader and more highly damped modes, numerical stability is reduced, and a more systematic analysis of convergence with respect to basis size, basis parameters, integration range, and scaling angle will be necessary.
In this sense, the present work should be viewed as a first systematic demonstration of the method rather than as its final optimized form.

\revone{An important future direction is the extension of the present construction to the separated Teukolsky equations for Kerr black holes \cite{Teukolsky:1972my, Leaver:1985ax}. This would test the applicability of CSM to rotating-black-hole ringdown and to a nonlinear spectral problem in which the radial equation depends on a frequency-dependent angular separation constant.}


\revone{An additional structural advantage of the present CSM framework is that the
continuum sector is encoded in the same complex-scaled spectral problem as
the isolated QNM poles. The corresponding continuum level density contains
information on the nonresonant background and the scattering-phase response
that is not present in a list of QNM frequencies alone. Appendix~\ref{app:cld} provides
only a proof-of-principle illustration. A more systematic CSM analysis of
the continuum response in de~Sitter black-hole backgrounds, including its
relation to the transmission phase and greybody factors, has been developed
in our subsequent work~\cite{Ogawa:2026opj}. The reconstruction of directly
observable scattering quantities remains an important direction for future
study.}

More broadly, our aim is to develop a formulation of black-hole QNMs that
remains spectrally transparent as one moves from standard benchmark problems
to more delicate backgrounds.
From this perspective, the complex scaling method offers a promising route
toward a unified treatment of resonance spectra in black-hole physics.

\section*{Acknowledgements}
We would like to thank Jiro Soda for helpful comments that inspired the idea for this work.
This work was partially supported by Japan Society for the Promotion of Science (JSPS) Grant-in-Aid for Scientific Research Grant Numbers JP25K17402 (O.M.), and 25H01267 (S.O.).
O.M.\ acknowledges the RIKEN Special Postdoctoral Researcher Program
and RIKEN FY2025 Incentive Research Projects.

\appendix
\section{Leaver's method and benchmark tables}\label{sec:leaver}
For later comparison, we briefly summarize Leaver's method for computing
quasinormal mode frequencies given in Refs.~\cite{Leaver:1985ax,Leaver:1986gd}.

Let $\psi_{l}(r)$ denote the radial wave function.
For the Regge--Wheeler equation, the effective potential vanishes both at the horizon and at spatial infinity.
Hence the quasinormal-mode boundary conditions are
\begin{align}
 \psi_{l}(r)\big|_{\mathrm{in}}
 &\sim e^{- \rmi \omega r_*}
 \sim (r-2M)^{-2 \rmi M \omega}
 \qquad
 r \to 2M,
\end{align}
and
\begin{align}
 \psi_{l}(r)\big|_{\mathrm{out}}
 &\sim e^{+ \rmi \omega r_*}
 \sim e^{\rmi \omega r} \, r^{2 \rmi M \omega}
 \qquad
 r \to \infty,
\end{align}
where $M$ is the black-hole mass, $\omega$ is the complex quasinormal frequency, and $r_*$ is the tortoise coordinate.

A convenient starting point is the generalized radial equation
\begin{align}
 \left[\Delta^{-s}\frac{\rmd}{\rmd r}\left(\Delta^{s+1}\frac{\rmd}{\rmd r}\right)-V(r)\right]
 \psi_{l}(r)
 =
 0,
 \label{eq:app-generalized-radial}
\end{align}
where $s$ is the spin weight, $V(r)$ is the radial potential, and
\begin{align}
\Delta = r^2 - 2 M r + a^2 .
\end{align}
Here $a$ is the Kerr rotation parameter.
The Schwarzschild case is recovered by setting $a=0$.

To implement the boundary conditions, one introduces the series ansatz
\begin{align}
 \psi_{l}(r)
 &= e^{\rmi \omega r}
 (r-r_-)^{-1-s+2 \rmi M \omega + 2 \rmi M \sigma}
 (r-r_+)^{-s-2 \rmi M \sigma}
 \sum_{n=0}^{\infty} a_n
 \left( \frac{r-r_+}{r-r_-} \right)^n ,
 \label{eq:app-leaver-ansatz}
\end{align}
with
\begin{align}
 r_{\pm} &= M \pm \sqrt{M^2-a^2},
 &
 \sigma &= \frac{1}{r_+ - r_-} \left( r_+ \omega - \frac{m a}{2M} \right).
\end{align}
Here $m$ is the azimuthal quantum number.
Substituting Eq.~\eqref{eq:app-leaver-ansatz} into Eq.~\eqref{eq:app-generalized-radial} yields a three-term recurrence relation for the expansion coefficients
$a_n$:
\begin{align}
 \alpha_0 a_1 + \beta_0 a_0 &= 0,
 \\
 \alpha_n a_{n+1} + \beta_n a_n + \gamma_n a_{n-1} &= 0,
 \qquad n \ge 1.
\end{align}
The coefficients $\alpha_n$, $\beta_n$, and $\gamma_n$ depend on the mode parameters,
in particular on $a$, $m$, and~$\omega$.

It is useful to define the ratio
\begin{align}
 A_n \equiv - \frac{a_n}{a_{n-1}} .
\end{align}
Then the recurrence relation implies
\begin{align}
 A_n = \frac{\gamma_n}{\beta_n - \alpha_n A_{n+1}} .
\end{align}
Iterating this relation gives the continued-fraction equation
\begin{align}
 \beta_0 &= \alpha_0 A_1
 = \frac{\alpha_0 \gamma_1}{
 \beta_1 - \frac{\alpha_1 \gamma_2}{
 \beta_2 - \frac{\alpha_2 \gamma_3}{
 \beta_3 - \cdots
 }}}.
 \label{eq:app-leaver-cf}
\end{align}

The quasinormal frequencies are obtained by solving Eq.~\eqref{eq:app-leaver-cf}.
In practice, one truncates the continued fraction at sufficiently large order.
Leaver solved, for example, the cases $l=2,3$ up to $N=60$ as in Table~\ref{tab:leaver_schwarzschild}.
Nollert later improved the large-$N$ convergence~\cite{Nollert:1993zz}.

Table~\ref{tab:onozawa_table1} summarizes QNM frequencies for extremal Reissner--Nordstr\"om black hole given by Onozawa \textit{et al.\/} in Ref.~\cite{Onozawa:1995vu}.
\revone{We follow the overtone-index conventions used in the corresponding benchmark literature. Accordingly, the Schwarzschild modes are labeled with the fundamental mode at $n=1$, whereas the extremal Reissner--Nordstr\"om modes, following Onozawa \textit{et al.\/}, are labeled with the fundamental mode at $n=0$. This difference is purely notational and does not affect the ordering of the modes.}

\begin{table}[ht]
\centering
\caption{$l=2$ and $l=3$ Schwarzschild gravitational QNMs by Leaver's method~\cite{Leaver:1985ax,Leaver:1986gd}.
$(\re\omega, \im\omega)$ is shown. Here $s=2$ and $M=0.5$.}
\label{tab:leaver_schwarzschild}
\begin{tabular}{rcc}
\toprule
$n$ & $l=2:\ \omega_n$ & $l=3:\ \omega_n$ \\
\midrule
 1 & $( 0.747343,\,-0.177925)$ & $( 1.198887,\,-0.185406)$ \\
 2 & $( 0.693422,\,-0.547830)$ & $( 1.165288,\,-0.562596)$ \\
 3 & $( 0.602107,\,-0.956554)$ & $( 1.103370,\,-0.958186)$ \\
 4 & $( 0.503010,\,-1.410296)$ & $( 1.023924,\,-1.380674)$ \\
 5 & $( 0.415029,\,-1.893690)$ & $( 0.940348,\,-1.831299)$ \\
 6 & $( 0.338599,\,-2.391216)$ & $( 0.862773,\,-2.304303)$ \\
 7 & $( 0.266505,\,-2.895822)$ & $( 0.795319,\,-2.791824)$ \\
 8 & $( 0.185617,\,-3.407676)$ & $( 0.737985,\,-3.287689)$ \\
 9 & $( 0.000000,\,-3.998000)$ & $( 0.689237,\,-3.788066)$ \\
10 & $( 0.126527,\,-4.605289)$ & $( 0.647366,\,-4.290798)$ \\
11 & $( 0.153107,\,-5.121653)$ & $( 0.610922,\,-4.794709)$ \\
12 & $( 0.165196,\,-5.630885)$ & $( 0.578768,\,-5.299159)$ \\
20 & $( 0.175608,\,-9.660879)$ & $( 0.404157,\,-9.333121)$ \\
30 & $( 0.165814,\,-14.677118)$ & $( 0.257431,\,-14.363580)$ \\
40 & $( 0.156368,\,-19.684873)$ & $( 0.075298,\,-19.415545)$ \\
41 & $( 0.154912,\,-20.188298)$ & $(-0.000259,\,-20.015653)$ \\
42 & $( 0.156392,\,-20.685530)$ & $( 0.017662,\,-20.566075)$ \\
50 & $( 0.151216,\,-24.693716)$ & $( 0.134153,\,-24.119329)$ \\
60 & $( 0.148484,\,-29.696417)$ & $( 0.163614,\,-29.135345)$ \\
\bottomrule
\end{tabular}
\end{table}

\begin{table}[ht]
\centering
\caption{Quasinormal frequencies for the extremal Reissner--Nordstr\"om black hole with $M=1$. The values $(\im \omega, \re \omega)$ are shown for each $s$ and~$l$. This table is taken from Ref.~\cite{Onozawa:1995vu} (Onozawa \textit{et al.\/}).}
\label{tab:onozawa_table1}
\begin{tabular}{clccc}
\toprule
$n$ & Method & $(s,l)=(2,2)$ & $(2,3)$ & $(2,4)$ \\
\midrule
0 & Onozawa& $(-0.083460,\,0.43134 )$ & $(-0.085973,\,0.70430 )$ 
           & $(-0.087001,\,0.96576 )$ \\
  & Leaver & $(-0.083645,\,0.43098 )$ & ---                     
           & ---                      \\
  & WKB    & $(-0.08349 ,\,0.43013 )$ & $(-0.08596 ,\,0.70398 )$ 
           & $(-0.08700 ,\,0.96563 )$ \\
\addlinespace
1 &  Onozawa& $(-0.25498 ,\,0.40452 )$ & $(-0.25992 ,\,0.68804 )$ 
           & $(-0.26212 ,\,0.95381 )$ \\
  & Leaver & $(-0.257055,\,0.39309 )$ & ---                     
           & ---                      \\
  & WKB    & $(-0.25675 ,\,0.40076 )$ & $(-0.26014 ,\,0.68701 )$ 
           & $(-0.26218 ,\,0.95339 )$ \\
\addlinespace
2 & Onozawa& $(-0.44137 ,\,0.35340 )$ & $(-0.44007 ,\,0.65624 )$ 
           & $(-0.44064 ,\,0.93020 )$ \\
  & Leaver & $(-0.442035,\,0.353515)$ & ---                     
           & ---                      \\
  & WKB    & $(-0.44210 ,\,0.35136 )$ & $(-0.43986 ,\,0.65575 )$ 
           & $(-0.44056 ,\,0.93004 )$ \\
\midrule
$n$ & Method & $(s,l)=(1,1)$ & $(1,2)$ & $(1,3)$ \\
\midrule
0 & Onozawa& $(-0.083460,\,0.43134 )$ & $(-0.085973,\,0.70430 )$ 
           & $(-0.087001,\,0.96576 )$ \\
  & Leaver & $(-0.08343 ,\,0.431415)$ & $(-0.086205,\,0.704075)$
           & ---                      \\
  & WKB    & $(-0.08349 ,\,0.43013 )$ & $(-0.08596 ,\,0.70398 )$ 
           & $(-0.08700 ,\,0.96563 )$ \\
\addlinespace
1 & Onozawa& $(-0.25498 ,\,0.40452 )$ & $(-0.25992 ,\,0.68804 )$  
           & $(-0.26212 ,\,0.95381 )$ \\
  & Leaver & $(-0.259705,\,0.40602 )$ & $(-0.26256 ,\,0.68315 )$ 
           & ---                      \\
  & WKB    & $(-0.25675 ,\,0.40076 )$ & $(-0.26014 ,\,0.68701 )$ 
           & $(-0.26218 ,\,0.95339 )$ \\
\addlinespace
2 & Onozawa& $(-0.44137 ,\,0.35340 )$ & $(-0.44007 ,\,0.65624 )$  
           & $(-0.44064 ,\,0.93020 )$ \\
  & Leaver & $(-0.44260 ,\,0.35347 )$ & $(-0.4408 ,\,0.655675 )$
           & ---                      \\
  & WKB    & $(-0.44210 ,\,0.35136 )$ & $(-0.43986,\,0.65575  )$ 
           & $(-0.44056 ,\,0.93004 )$ \\
\midrule
$n$ & Method & $(s,l)=(0,0)$ & $(0,1)$ & $(0,2)$ \\
\midrule
0 & Onozawa& $(-0.095844,\,0.13346 )$ & $(-0.089384,\,0.37764 )$
           & $(-0.088748,\,0.62657 )$ \\
  & WKB    & $(-0.10371 ,\,0.12109 )$ & $(-0.08936 ,\,0.37570 )$
           & $(-0.08873 ,\,0.62609 )$ \\
\addlinespace
1 & Onozawa& $(-0.33065 ,\,0.092965)$ & $(-0.27614 ,\,0.34818 )$
           & $(-0.26909 ,\,0.60817 )$ \\
  & WKB    & $(-0.33742 ,\,0.09157 )$ & $(-0.27828 ,\,0.34392 )$
           & $(-0.26944 ,\,0.60677 )$ \\
\addlinespace
2 & Onozawa& $(-0.58833 ,\,0.075081)$ & $(-0.48643 ,\,0.29846 )$ 
           & $(-0.45820 ,\,0.57287 )$ \\
  & WKB    & $(-0.57164 ,\,0.05056 )$ & $(-0.48145 ,\,0.29661 )$
           & $(-0.45750 ,\,0.57254 )$ \\
\bottomrule
\end{tabular}
\end{table}

\section{A note on continuum level density in the complex-scaling framework}
\label{app:cld}

In the main text, we have focused on the extraction of quasinormal-mode (QNM)
frequencies as isolated resonance eigenvalues of the complex-scaled operator.
For the present paper, this is the primary use of the complex scaling method
(CSM).
However, the CSM is not limited to discrete resonance poles.
It also provides a natural framework for treating the continuum sector, which
is physically important in black-hole perturbation theory.

From the black-hole point of view, this issue is closely related to the
late-time tail.
While the intermediate-time ringdown is governed by QNM poles, the late-time
signal is associated with continuum contributions, often described in terms of
a branch cut of the Green function and the resulting power-law decay.
For this reason, a quantity that probes the continuum sector is of direct
physical interest.
In the CSM framework, such a quantity is the continuum level density (CLD)~\cite{Suzuki:2005wv}.

Let $H$ be the full operator and $H_0$ a suitable reference operator.
The CLD is formally defined by
\begin{align}
    \Delta \rho(E)
    =
    -\frac{1}{\pi}
    \im
    \Tr
    \left[
        \frac{1}{E-H+\rmi 0}
        -
        \frac{1}{E-H_0+\rmi 0}
    \right],
    \label{eq:cld_def_appendix}
\end{align}
where $E$ is the spectral parameter~\cite{Avishai1985}.
After complex scaling, this becomes
\begin{align}
    \Delta \rho^\theta(E)
    &=
    -\frac{1}{\pi}
    \im
    \Tr
    \left[
        \frac{1}{E-H^\theta}
        -
        \frac{1}{E-H_0^\theta}
    \right]
    \\
    &=
    -\frac{1}{\pi}
    \im
    \left[
        \sum_\nu\frac{1}{E-E^\theta_\nu}
        -
        \sum_{\nu'}\frac{1}{E-E_{0,\nu'}^\theta}
    \right],
    \label{eq:cld_csm_appendix}
\end{align}
where $E^\theta_\nu$ and $E_{0,\nu'}^\theta$ are complex energies of complex-scaled Hamiltonians $H^\theta$ and $H^\theta_0$, respectively.
The point is that, in the CSM, both isolated resonances and the rotated
continuum are represented within the same non-Hermitian spectral problem.
Therefore, the same framework that extracts QNM frequencies can, in principle,
also be used to analyze continuum observables relevant to late-time tails.
Another important point should be emphasized. As seen from Eq.~\eqref{eq:cld_csm_appendix}, the CLD is expressed as an incoherent sum of terms of the form $1/(E-E^\theta_\nu)$. This means that, because each resonance is completely isolated as a single eigenstate in the CSM, its contribution to the CLD can be extracted separately.

It is helpful to explain the meaning of this expression in simple terms.
If one considers only discrete bound states, then the spectrum consists of
isolated energy levels.
By contrast, in a scattering problem the spectrum also contains a continuum.
The CLD describes how this continuum is modified by the interaction.
In other words, it is not merely a count of isolated states, but a spectral
measure of how the interacting system differs from the free or reference one
throughout the continuum region.

In standard scattering theory, the CLD is closely related to phase-shift
information.
For a single channel, one may schematically write
\begin{align}
    \Delta \rho(E)
    \propto
    \frac{d\delta(E)}{dE},
\end{align}
where $\delta(E)$ is the scattering phase shift.
This relation is well known in other areas of physics and is one of the
reasons why the CLD is a useful bridge between spectral theory and scattering
observables.
\revone{The physical information contained in the CLD is complementary to, but
different from, a list of QNM frequencies. The QNM frequencies characterize
the isolated pole sector, whereas the CLD also retains the nonresonant
continuum background and describes how the continuum density is modified
by the black-hole potential. Through its relation to the derivative of the
scattering or transmission phase, it provides information on the continuum
phase response that is absent from the QNM spectrum alone.}

\revone{The CLD should not, however, be identified directly with an observable
transmission probability or a greybody factor. It characterizes the
continuum density and phase response, whereas a greybody factor also
requires the modulus of the transmission amplitude. A systematic analysis
of the continuum response, including its relation to the transmission phase
and greybody factors, is presented in our subsequent work \cite{Ogawa:2026opj}.
The purpose of the present appendix is limited to demonstrating that the
same finite-basis CSM framework can access both the pole and continuum
sectors.}

In the present work, we do not attempt a systematic CLD analysis for all
backgrounds.
A full study would require a careful choice of the reference operator $H_0$,
together with a dedicated investigation of the dependence on the scaling
angle, basis truncation, and numerical integration range.
Nevertheless, it is useful to record that the present CSM formulation is in
principle capable of treating not only resonance poles but also the continuum
sector.

As a simple illustration, one may consider the extremal
Reissner--Nordstr\"om case in a channel where the QNM calculation is
numerically most stable, for example the odd-parity electromagnetic mode with
$(s,l)=(1,3)$ or equivalently the odd-parity gravitational mode with
$(s,l)=(2,4)$.
In such a case, the low-lying resonance poles are already well separated and
accurately reproduced in the main analysis, so the same setup is a natural
starting point for a CLD plot.
In the present formulation, the spectral variable of the Schr\"odinger-type equation is $E=\omega^2$. Therefore, when we express the CLD in terms of $\omega$ rather than $E$, the Jacobian of the transformation should be taken into account as
\begin{align}
    \Delta^{\theta}(\omega)=\frac{dE}{d\omega}\Delta^{\theta}(E)
    = 2\omega\Delta^{\theta}(E).
\end{align}

\begin{figure}[ht]
    \centering
    \includegraphics[width=0.48\columnwidth]{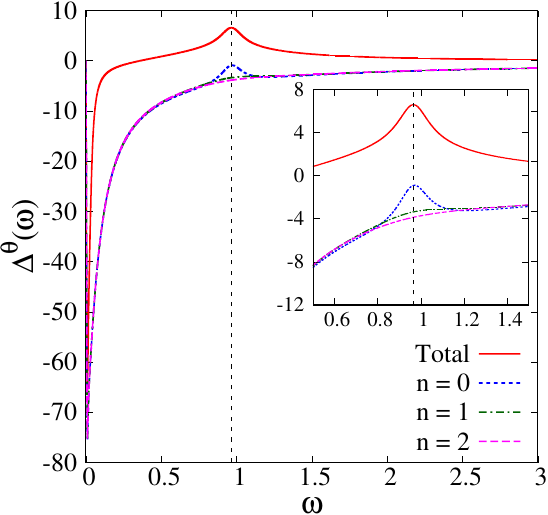}
    \hspace{1em}
    \includegraphics[bb=0 0 270 252,clip,width=0.45\columnwidth]{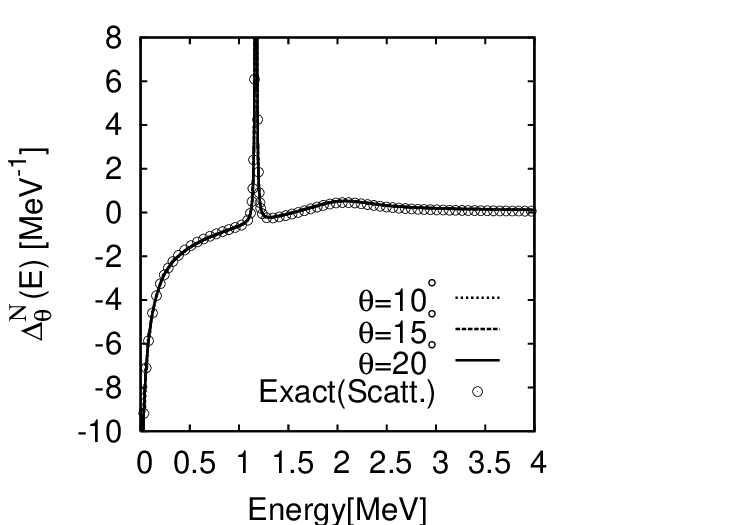}
    \caption{\textbf{Left panel:} Illustrative CLD in the CSM for the extremal Reissner--Nordstr\"om case in the odd-parity electromagnetic channel $(s,l)=(1,3)$, together with its decomposition into contributions from the low-lying modes $\omega_{n=0,1,2}$.
    We set $H_0=-\rmd^2/\rmd r_*^2$ and $\theta=40^\circ$. The vertical dashed line represents the value of $\re \omega_{\revone{0}}$.
    The CLD shows a characteristic structure: it is negative on the low-energy side, develops a pronounced peak near the resonance energy, and then gradually approaches zero at higher energies.
    This is qualitatively the same pattern as in standard CSM applications to resonance problems, indicating that the present black-hole calculation captures a resonance embedded in the continuum in essentially the same spectral manner.
    In black-hole language, this is the sector that is expected to be related to branch-cut contributions and late-time power-law tails.
    \textbf{Right panel:} A typical CLD in a nuclear system taken from Ref.~\cite{Suzuki:2005wv}, where one finds the same qualitative behavior: a negative contribution on the low-energy side, a resonance peak, and an asymptotic approach to zero.}
    \label{fig:cld_illustration}
\end{figure}

A representative result is shown in
Fig.~\ref{fig:cld_illustration} as a function of $\omega$
(or equivalently of $E$) with $\theta=40^\circ$.
The full CLD (the solid line) exhibits the characteristic resonance profile familiar in the CSM: it is negative on the low-energy side, develops a pronounced peak near the resonance region, and then gradually approaches zero at higher energies. Since the CSM allows one to separate the contributions from individual resonance states, we also show the decomposition into the contributions of the
lowest mode $\omega_0$ and of the higher modes $\omega_1$ and $\omega_2$, which are represented by the dotted, dot-dashed, and dashed curves, respectively.
The dominant peak structure is produced mainly by the lowest resonance
$\omega_0$, whose contribution already shows essentially the same qualitative behavior as the full CLD, namely a negative low-energy part, a resonance peak, and an asymptotic approach to zero.
By contrast, the contributions from $\omega_1$ and $\omega_2$ do not display
comparably visible isolated peaks in the present range, but rather yield a
smoother component that changes from a negative value toward zero.

\revone{Figure~\ref{fig:cld_illustration} should therefore be regarded as a proof-of-principle
illustration. It shows that the finite-basis CSM calculation used for the
QNM frequencies also contains information on the nonresonant continuum
response and permits a pole-by-pole decomposition of the CLD. A systematic
analysis of this continuum sector has been pursued in Ref.~\cite{Ogawa:2026opj},
while the reconstruction of directly observable scattering quantities
remains a subject for future work.}



\bibliographystyle{utphys}
\bibliography{ref,ref_ewkb,ref_res,ref_scattering}
\end{document}